\documentclass[showpacs,showkeys,preprintnumbers,pre,amsmath,amssymb,twocolumn]{revtex4-2}
\usepackage[english]{babel}
\usepackage[utf8]{inputenc}
\usepackage{amsthm}
\usepackage{color}
\usepackage{tikz-cd}

\usepackage[colorlinks=true,linkcolor=blue,urlcolor=blue,citecolor=blue]{hyperref}
\usepackage{bm}

\begin{document}

\title{Higher-order Tuning of Interface Physics in Multiphase Lattice Boltzmann}

\author{Matteo Lulli}
\email{mlulli@phy.cuhk.edu.hk}
\author{Emily S. C. Ching}
\affiliation{Department of Physics, The Chinese University of Hong Kong, Shatin, Hong Kong, China}

\begin{abstract}
Tuning the interface properties of multiphase models is of paramount importance to the final goal of achieving a one-to-one matching with nucleation and cavitation experiments. The surface tension, at the leading order, and the Tolman length, at higher order, play a crucial role in the estimation of the free-energy barrier determining the experimentally observed nucleation rates. The lattice Boltzmann method allows for a computationally efficient modelling approach of multiphase flows, however, tuning results are concerned with the surface tension and neglect the Tolman length. We present a novel perspective that leverages all the degrees of freedom hidden in the forcing stencil of the Shan-Chen multiphase model. By means of the lattice pressure tensor we determine and tune the coefficients of higher-order derivative terms related to surface tension and Tolman length at constant interface width and density ratio. We test the method by means of both hydrostatic and dynamic simulations and demonstrate the dependence of homogeneous nucleation rates on the value of the Tolman length. This work provides a new tool that can be integrated with previously existing strategies thus marking a step forwards to a high-fidelity modelling of phase-changing fluid dynamics.
\end{abstract}

\maketitle

\section{Introduction}\label{sec:intro}
The hydrodynamics of phase-changing fluids is at the foundation of nowadays major challenges both from an engineering and a theoretical perspective. In this arena, accurate numerical simulations would provide an important tool for the development of more sophisticated physical modelling of out-of-equilibrium multiphase interface physics problems that are extremely consequential for industrial applications. Nucleation~\cite{Debenedetti1996,Kalikmanov2013,LANDAU1992,LIFSHITZ1980}, cavitation and jet-breakup dynamics~\cite{brennen_1995_book_cav_bub,Chryssakis_2006,Ohl2006} all belong to this class of problems with applications ranging from the development of more efficient designs for hydrolysis cells to that of propeller blades or engines injectors, as well as of nozzles for large-scale industrial painting processes. In all the above processes there is a non-trivial coupling between the confined hydrodynamics and the multiphase physics yielding to phase transformation as a response to thermal fluctuations and/or to hydrodynamic stresses.

From the physics perspective, the relevant parameters are (i) the density ratio between liquid and gas phases, (ii) the surface tension associated with a flat equilibrium interface and (iii) the curvature corrections to the surface tension for closed interfaces. With these quantities it is possible to estimate nucleation rates by means of Classical Nucleation Theory (CNT): given that nucleation is an activated process, its rate $J$ is proportional to $\exp[-\Delta F / k_B T]$, where $k_B$ is the Boltzmann constant, $T$ the temperature and $\Delta F$ the free-energy barrier. In the \emph{capillary approximation}, i.e., when using the flat-interface surface tension $\sigma_0$, one can write $\Delta F\propto-4\pi R_{c}^{2}\sigma_{0}$~\cite{Kalikmanov2013} where $R_c$ is the critical radius above which the nucleating cluster would keep on growing. However, CNT does not always provide good estimates of nucleation rates~\cite{Debenedetti1996,Kalikmanov2013} and may also lead to inconsistent predictions, such as critical clusters with a negative number of molecules~\cite{Aasen2020}. Indeed, the critical size $R_c$ is not uniquely defined, especially when considering the interface as a smooth, although steep, change of density. Gibbs~\cite{GibbsCollected1948} introduced the concept of \emph{arbitrary dividing surface} whose curvature explicitly appears in the free-energy, which, at equilibrium, is expected to be independent of its value, i.e., stationary. In the case of curved interfaces one obtains a generalized surface tension $\sigma[R]$ that can be used to write a generalized Laplace law augmented by the term $\mbox{d}\sigma[R]/\mbox{d}R$
\begin{equation}
    \Delta P = \frac{(d-1)\sigma[R_s]}{R_s} + \frac{\mbox{d}\sigma[R]}{\mbox{d}R}\bigg|_{R = R_s} = \frac{(d-1)\sigma(R_s)}{R_s},
    \label{eq:ExactLaplace}
\end{equation}
where $d$ is the space dimension and $R_s$, the position of the \emph{surface of tension}, is defined by the vanishing of the derivative $\mbox{d}\sigma[R]/\mbox{d}R|_{R_s} = 0$. The surface tension $\sigma$ will in general retain a dependence on the bubble/droplet radius $R_s$ which can be parametrized as a power-law expansion in the curvature $R_s^{-1}$
\begin{equation}
\sigma\left(R_{s}\right)\simeq\sigma_{0}\left[1-\frac{(d-1)\delta}{R_{s}}\right].
\label{eq:SigmaCExpansion}
\end{equation}
The leading order contribution $\sigma_0$ is the flat-interface value, the first-order coefficient $\delta$ is the so-called Tolman length~\cite{Tolman1949}. Considering Eq.~\eqref{eq:SigmaCExpansion}, one can extend CNT changing the size-dependence of $\Delta F$: although corrections may be small, there can be sizeable changes in the rates estimation due to their exponential dependence on $\Delta F$. The importance of curvature corrections for extending CNT has been investigated in several experimental and theoretical works~\cite{Talanquer1995,Tanaka_2015,Bruot2016,Nguyen2018,Aasen2020}.

Density functional theory has been used to provide an analytical expression for the Tolman length for multicomponent and multiphase mixtures~\cite{Aasen2018,Rehner2018,Blokhuis2013,Wilhelmsen2015}. Nevertheless, such studies cannot easily be extended to include hydrodynamics, hence, numerical modelling plays a crucial role. On one hand, Molecular Dynamics (MD) offers an invaluable tool for understanding non-equilibrium phenomena, especially at short time and length scales~\cite{VanGiessen2009,Nijmeijer1992,Schrader2009,Menzl2016}, yielding results that are implicitly compliant, from the ground up, with the microscopic foundation of thermodynamics and hence with experiments. However, atomistic modelling comes with a very large computational cost for simulating systems in the hydrodynamic limit. On the other hand, as the scale grows, mesoscopic approaches, like explicit methods and kinetic models, successfully came into play~\cite{Anderson_1998,magaletti2016shockinduced,Gallo2018,Gallo2020}. However, no numerical approach has been able to controllably tune the Tolman length so far.

The lattice Boltzmann method (LBM)~\cite{Kruger_2017,succi2018lattice} uses a discrete version of the Boltzmann equation recovering Navier-Stokes through a multi-scale (Chapman-Enskog) expansion. Several approaches for multiphase modelling are available in LBM, namely, free-energy-based approaches~\cite{Swift_1995,Swift_1996,Guo_2021,Hosseini_2022}, color-gradient~\cite{Gunstensen_1991,Montessori_2019}, phase-field~\cite{he1999a,fakhari2010phasefield} and pseudo-potential approaches~\cite{Shan_1993,Shan_1994}. The last class, typically referred to as Shan-Chen (SC) model, is characterized by the use of a pseudo-potential function $\psi$ instead of the density $n$ when computing the body force used to obtain a non-ideal equation of state. The SC model is also equipped with a \emph{lattice pressure tensor}~\cite{Shan_2008,Sbragaglia_2013,Lulli_2021} which is central for the determination of the Tolman length~\cite{Lulli_2022,lulli2022mesoscale} and for the modelling of hydrodynamic fluctuations~\cite{lulli2024metastable}.

This paper presents the first computational results allowing for a quantitative control of the Tolman length $\delta$ in the context of the SC multiphase model. This marks an important progress with respect to the more phenomenological approaches adopted in~\cite{Menzl2016,Gallo2018,Lulli_2022,lulli2022mesoscale,Hosseini_2022} where the value of $\delta$ was only estimated \emph{a posteriori}. We derive an analytical expression for $\delta$ by leveraging the one-dimensional projection of the lattice pressure tensor. This is obtained by expressing the interaction potential weights as a function of the stencil~\emph{moments}, i.e., as a function of the physical knobs. These results rely on the control of the interaction potential isotropy~\cite{Lulli_2021}, which, different from the existing literature, is not taken as maximal, thus allowing us to unlock the so-far hidden degrees of freedom. 

The paper is organized as follows: in Sec.~\ref{sec:lbm} we present the fundamentals of the lattice Boltzmann method (LBM) introducing some new notations that highlights the statistical interpretation of lattice quadratures; in Sec.~\ref{sec:SC} we briefly review the lay ideas of the Shan-Chen model; in Sec.~\ref{sec:met} we present our new approach for the higher-order tuning for surface tension and Tolman length; in Sec.~\ref{sec:results} we provide a new strategy for the control of spurious currents, report the results for the tuning of the interface properties in the hydrostatic case, and demonstrate the dependence of homogeneous nucleation rates on the value of the Tolman length; finally in Sec.~\ref{sec:dandc} we provide some discussion and perspectives. The source code for the simulations can be found on the github repository \href{https://github.com/lullimat/idea.deploy}{https://github.com/lullimat/idea.deploy}~\cite{sympy, scipy, numpy0, numpy1, scikit-learn, matplotlib, ipython, pycuda_opencl}, where a Jupyter notebook~\cite{ipython} is available to reproduce all the results reported in this paper.

\section{Lattice Boltzmann}\label{sec:lbm}
LBM is formulated as a discretized version of the Boltzmann transport equation: the spatial dependence $\mathbf{x}$ is defined on a regular lattice, the particle momentum $\boldsymbol{\xi}$ is also discretized in what is known as a lattice quadrature (or velocity stencil)~\cite{Kruger_2017,succi2018lattice,Shan_2016}, which is a set of weights $\{w_i\}$ and vectors $\{\boldsymbol{\xi}_i\}$ respecting certain isotropy conditions (see Appendix~\ref{app:e4ex}) and which connect each point of the lattice to its neighbors. The discrete time is provided by the alternation of two basic update steps known as \emph{collision} and \emph{streaming}: in the former the discrete distribution is updated \emph{locally} through the use of a collision operator $\Omega$ which exactly conserves mass and momentum, while in the latter the updated distribution is moved to the adjacent nodes. Considering the discrete dependence of probability density $f$ on the particle momentum, one defines the quantities $f_i=f(\mathbf{x}, \boldsymbol{\xi}_i, t)$, the so-called \emph{populations}. We introduce the notation $\langle g(x)\rangle_\omega$ which is typically used in statistical mechanics, to indicate the expectation value of a function of a random variable $g(x)$ with respect to a distribution $\omega$, be it continuous or discrete. Using this notation density and momentum can be expressed as
\begin{equation}
n=\sum_{i}f_{i}=\langle1\rangle_{f},\qquad nu^{\mu}=\sum_{i}f_{i}\xi_{i}^{\mu}=\langle\xi^{\mu}\rangle_{f}.
\end{equation}
We extend the same notation to indicate the finite sum of a function over a set of weights of arbitrary sign. The lattice Boltzmann equation reads
\begin{equation}\label{eq:lbm}
	f_{i}\left(\mathbf{x}+\boldsymbol{\xi}_{i},t+1\right)-f_{i}\left(\mathbf{x},t\right)=\Omega_{i}\left(\mathbf{x},t\right)+F_{i}\left(\textbf{x},t\right)
\end{equation}
where the r.h.s. represents the collision step while the l.h.s. the streaming one. The simplest collision operator $\Omega_i$ is the single-relaxation-time Bhatnagar, Gross and Krook (BGK) operator~\cite{Bhatnagar_1954}
\begin{equation}\label{eq:omega_i}
    \Omega_{i}\left(\mathbf{x},t\right)=-\frac{1}{\tau}\left[f_{i}\left(\mathbf{x},t\right)-f_{i}^{\left(\text{eq}\right)}\left(\mathbf{x},t\right)\right]
\end{equation}
which relaxes all populations towards an equilibrium distribution $f_{i}^{\left(\text{eq}\right)}$ with a rate $\tau^{-1}$. The equilibrium distribution is an approximation of the Maxwell-Boltzmann distribution, i.e., a Gaussian. The expression for $f_{i}^{\left(\text{eq}\right)}$ is obtained via a Taylor expansion on the Hermite polynomial basis to second order~\cite{SHAN_2006,Kruger_2017,succi2018lattice}
\begin{equation}\label{eq:f_eq}
\begin{split}
&f_{i}^{\left(\text{eq}\right)}=w_{i}n\\&\quad\times\left[1+\frac{1}{c_{s}^{2}}\xi_{i}^{\alpha}u_{\alpha}^{\left(\text{eq}\right)}+\frac{1}{2c_{s}^{4}}\left(\xi_{i}^{\alpha}\xi_{i}^{\beta}-c_{s}^{2}\delta^{\alpha\beta}\right)u_{\alpha}^{\left(\text{eq}\right)}u_{\beta}^{\left(\text{eq}\right)}\right]\end{split}
\end{equation}
where $\mbox{He}_0(\xi)=1$, $\mbox{He}^\mu_1(\xi)=\xi^\mu/c_s^2$ and $\mbox{He}^{\mu\nu}_2(\xi)=(\xi^\mu\xi^\nu - c_s^2\delta^{\mu\nu})/c_s^4$, with $c_s^2$ being the square of the speed of sound. The stencil used in Eq.~\eqref{eq:f_eq} is chosen to be a lattice Gaussian quadrature~\cite{Shan_2016} with variance $c_s^2$, indicated with the notation $DdQq$ with $d$ the dimension and $q$ the number of vectors. The most common choices for two and three-dimensions are $D2Q9$ and $D3Q19$ which ensure isotropy up to the fourth moment, i.e.,
\begin{equation}
\langle1\rangle_{w}=1,\quad\langle\xi^{\mu}\xi^{\nu}\rangle_{w}=c_{s}^{2}\delta^{\mu\nu},\quad\langle\xi^{\mu}\xi^{\nu}\xi^{\rho}\xi^{\sigma}\rangle_{w}=c_{s}^{4}\Delta^{\mu\nu\rho\sigma},
\end{equation}
where $\Delta^{\mu\nu\rho\sigma}=\delta^{\mu\nu}\delta^{\rho\sigma} + \delta^{\mu\rho}\delta^{\nu\sigma} + \delta^{\mu\sigma}\delta^{\nu\rho}$ is the rank-four fully isotropic tensor. In the case $n\mathbf{u}=n\mathbf{u}^{\text{eq}}$, the first two moments of the equilibrium distribution coincide with the population ones, i.e.,
\begin{equation}
n=\langle1\rangle_{\text{eq}}=\langle1\rangle_{f},\quad nu^{\mu}=\langle\xi^{\mu}\rangle_{\text{eq}}=\langle\xi^{\mu}\rangle_{f},
\end{equation}
and from this fact one can immediately see that the collision operator exactly conserves mass and momentum
\begin{equation}
\langle1\rangle_{\Omega}=\langle\xi^{\mu}\rangle_{\Omega}=0,
\end{equation}
which is a necessary condition to recover Navier-Stokes equations. The term $F_i(\mathbf{x}, t)$ in Eq.~\eqref{eq:lbm} is used to introduce a body force: we follow the construction proposed by Guo~\cite{Guo_2002} that is also written in terms of a Hermite-polynomials expansion
\begin{equation}
\begin{split}F_{i}\left(\mathbf{x},t\right) & =\left(1-\frac{1}{2\tau}\right)w_{i}\\
\times & \left[\frac{1}{c_{s}^{2}}\xi_{i}^{\alpha}F_{\alpha}+\frac{1}{2c_{s}^{4}}\left(\xi_{i}^{\alpha}\xi_{i}^{\beta}-c_{s}^{2}\delta^{\alpha\beta}\right)T_{\alpha\beta}\right],
\end{split}\label{eq:GuoForcing}
\end{equation}
where the rank-two tensor $T_{\alpha\beta}$ is typically chosen as $T_{\alpha\beta}=2u_{(\alpha} F_{\beta)}=2u_\alpha F_\beta$ (given that the second-degree Hermite polynomial is a symmetric tensor one can omit the explicit symmetrization in $T_{\alpha\beta}$). In order for the scheme to correctly recover hydrodynamics the equilibrium velocity must be \emph{shifted} by the force with respect to the raw populations momentum, i.e.,
\begin{equation}
u_{\left(\text{eq}\right)}^{\alpha}\left(\mathbf{x},t\right)=\frac{1}{n\left(\mathbf{x},t\right)}\langle\xi^{\alpha}\rangle_{f}+\frac{1}{2n\left(\mathbf{x},t\right)}F^{\alpha}\left(\mathbf{x},t\right),
\end{equation}
so that at second-order the multi-scale Chapman-Enskog expansion~\cite{Kruger_2017} recovers a specific version of the Navier-Stokes (NS) equations
\begin{equation}
\begin{split}\partial_{t}n+\partial_{\mu}\left(nu^{\mu}\right)=0\\
\partial_{t}\left(nu^{\mu}\right)+\partial_{\rho}\left(nu^{\rho}u^{\mu}\right)= & -\partial^{\mu}\left(nc_{s}^{2}\right)+F^{\mu}\\
 &+\partial_{\rho}\left[\mu_{s}\left(\partial^{\rho}u^{\mu}+\partial^{\mu}u^{\rho}\right)\right]
\end{split}\label{eq:NS}
\end{equation}
where $\mu_s = nc_s^2(\tau - 1/2)$ is the dynamic shear viscosity and $n c_s^2$ represent the ideal-gas equation of state so that $-\partial^\mu (n c_s^2)$ is the divergence of the ideal pressure tensor $P^{\mu\nu}=n c_s^2\delta^{\mu\nu}$. 

\section{Shan-Chen Model}\label{sec:SC}
\subsection{Bulk Physics}
\begin{figure*}[!t]
\includegraphics[scale=0.5]{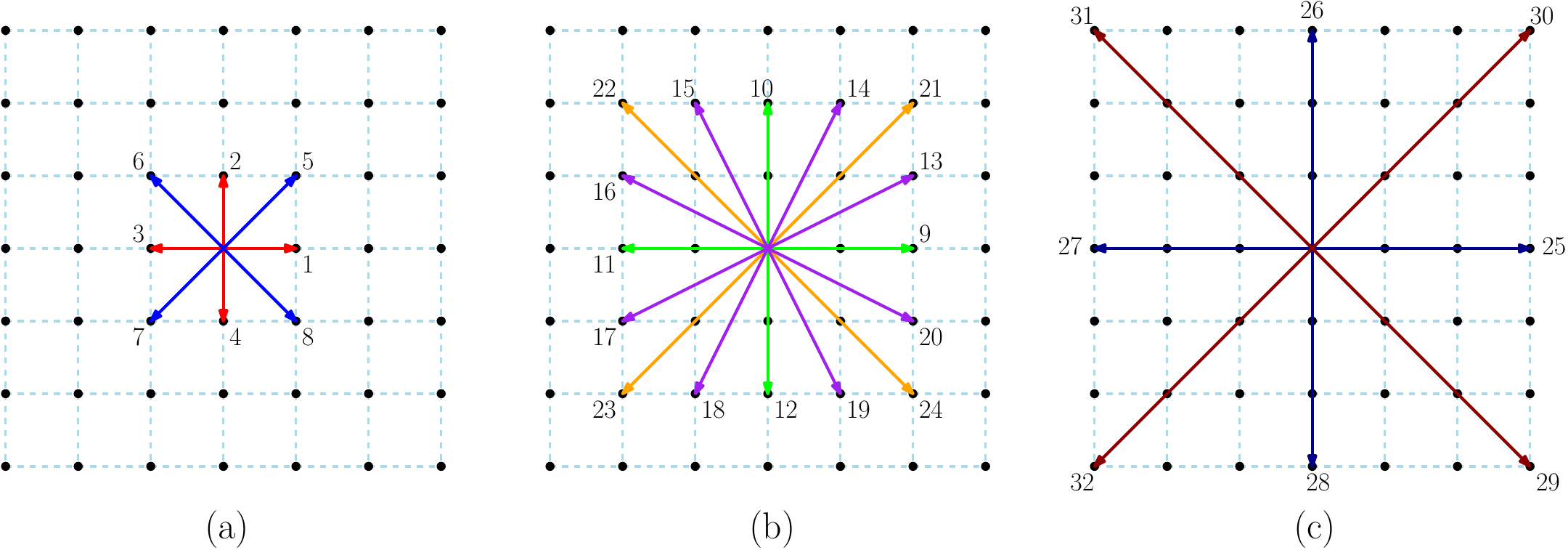}
\caption{Set of basis vectors $\{\mathbf{e}_a\}$ used to construct the forcing schemes presented for the two-dimensional simulations. Panels (a), (b) and (c) show the numbers referred to lattice vectors, with the color coding for the different squared lengths $\{|\mathbf{e}_{a}|^{2} = 1,2,4,5,8,9,18\}$.}\label{fig:stencil}
\end{figure*}
To introduce non-ideal interactions Shan and Chen~\cite{Shan_1993,Shan_1994} proposed the use of a body force of the form
\begin{equation}\label{eq:sc_f}
F^{\mu}\left(\mathbf{x}\right)=-Gc_{s}^{2}\psi\left(\mathbf{x}\right)\sum_{\mathbf{e}_{a}\in\mathcal{G}}W\left(|\mathbf{e}_{a}|^{2}\right)\psi\left(\mathbf{x}+\mathbf{e}_{a}\right)e_{a}^{\mu},
\end{equation}
where $G<0$ is a coupling constant and $\psi=\psi(n)$ is a local function of the density which needs to satisfy the saturation condition $\lim_{n\to\infty} \psi(n)=\psi_\infty$, with the common choice $\psi_\infty=1$. The forcing stencil, i.e., the set of weights $\{W(|\mathbf{e}_a|^2)\}$ and vectors $\{\mathbf{e}_a\}$ does not depend on the velocity stencil with weights $\{w_i\}$ and abscissas $\{\boldsymbol{\xi}_i\}$. However, one can still understand the $W$'s and the $\mathbf{e}$'s to represent a lattice quadrature of an arbitrary scalar function. Hence, by using the expectation value notation, we can rewrite Eq.~\eqref{eq:sc_f} using a more compact expression
\begin{equation}
F^{\mu}\left(\mathbf{x}\right)=-Gc_{s}^{2}\psi\left(\mathbf{x}\right)\langle\psi\left(\mathbf{x}+\mathbf{e}\right)e^{\mu}\rangle_{W}.
\end{equation}

In order to capture the isotropy of the bulk physics, both the velocity and the forcing stencil vectors are always chosen in a symmetric way (see Fig.~\ref{fig:stencil}), i.e., vectors belong to \emph{dihedral} groups where each element can be transformed into another by the operation of coordinates permutations and sign flip. It also follows that for each vector belonging to a given symmetry group the associated weight assumes always the same value. Hence, only the expectation values of an \emph{even} number of vectors would be non-vanishing and the Taylor expansion of the forcing reads
\begin{equation}
\begin{split}
F^{\mu}\left(\mathbf{x}\right)=&-Gc_{s}^{2}\psi\left(\mathbf{x}\right)\times\\
\times&\sum_{k=0}^{+\infty}\frac{1}{(2k + 1)!}\partial_{\nu_{1}\ldots\nu_{2k+1}}\psi\left(\mathbf{x}\right)\langle e^{\nu_{1}}\cdots e^{\nu_{2k+1}}e^{\mu}\rangle_{W}.\label{eq:FSCTaylor}
\end{split}
\end{equation}
At the leading order, the force is proportional to the first derivative of $\psi$ contracted with the second moment
\begin{equation}
F^{\mu}\left(\mathbf{x}\right)\simeq-\partial_{\nu}\left[\frac{Gc_{s}^{2}e_{2}}{2}\psi^{2}\left(\mathbf{x}\right)\delta^{\mu\nu}\right].\label{eq:FSCLeading}
\end{equation}
since the second moment of the forcing stencil is necessarily proportional to the Kronecker delta, which is an isotropic tensor, i.e., $\langle e^{\mu}e^{\nu}\rangle_{W}=e_{2}\delta^{\mu\nu}$.
Eq.~\eqref{eq:FSCLeading} can be inserted in the NS equations~\eqref{eq:NS} and summed to the ideal-gas contribution yielding the bulk pressure tensor
\begin{equation}
P_{b}^{\mu\nu}\left(\mathbf{x}\right)=\left[n\left(\mathbf{x}\right)c_{s}^{2}+\frac{Gc_{s}^{2}e_{2}}{2}\psi^{2}\left(\mathbf{x}\right)\right]\delta^{\mu\nu},
\end{equation}
which, for appropriate choices of $\psi$, can induce phase separation when the coupling constant is below some critical value $G_c$. The second moment of the forcing stencil is commonly set to $e_2=1$.

\subsection{Interface Physics}\label{ssec:interface}
We continue now by reintroducing the ideal gas term and projecting the lattice pressure tensor to one dimension in order to consider a flat interface profile with gradients parallel to the $x$ direction (see Appendix~\ref{app:LPT2D} for the full expressions in two and three dimensions). We set the normal as $P^{xx}(x) = P_\text{N}(x)$ and the tangential one as $P^{yy}(x) = P_\text{T}(x)$. Since the lattice pressure tensor exactly obeys the mechanic equilibrium condition $P_\text{N}(x)=p_0$ on the lattice~\cite{Sbragaglia_2013,Lulli_2021,lulli2024metastable}, $P_\text{N}(x)$ is a non-local function of the fields that can be understood as the exact summation of a series expansion, so that the truncation error in the Taylor expansion is well-behaved. The Taylor expansion of the normal component up to second order reads
\begin{equation}
\begin{split} & P_{N}\left(x\right)=n\left(x\right)c_{s}^{2}+\frac{Gc_{s}^{2}e_{2}}{2}\psi^{2}\left(x\right)\\
 & \quad+Gc_{s}^{2}\left\{ \bar{\alpha}\left[\frac{\text{d}\psi\left(x\right)}{\text{d}x}\right]^{2}+\bar{\beta}\psi\left(x\right)\frac{\text{d}^{2}\psi\left(x\right)}{\text{d}x^{2}}\right\},
\end{split}\label{eq:SCPNExpansion}
\end{equation}
where $\bar{\alpha}=\alpha/12$ and $\bar{\beta}=\beta/12$ are related to $\alpha$ and $\beta$ used in~\cite{Shan_2008,Lulli_2021}. By means of the identity $\frac{\text{d}^{2}\psi}{\text{d}x^{2}}=\frac{1}{2}\frac{\text{d}}{\text{d}\psi}\left[\frac{\text{d}\psi}{\text{d}x}\right]^{2}$ the terms inside the curly brackets can be written as a total derivative
\begin{equation}
\bar{\alpha}\left[\frac{\text{d}\psi}{\text{d}x}\right]^{2}+\bar{\beta}\psi\frac{\text{d}^{2}\psi}{\text{d}x^{2}}=\frac{\bar{\beta}}{2}\psi^{\varepsilon+1}\frac{\text{d}}{\text{d}\psi}\left\{ \psi^{-\varepsilon}\left[\frac{\text{d}\psi}{\text{d}x}\right]^{2}\right\},
\end{equation}
where $\varepsilon = -2\bar{\alpha}/\bar{\beta}$. We now express $\varepsilon$ in terms of the isotropy coefficients~\cite{Shan_2008,Lulli_2021} which results in a function of $e_4$ only (since we set $e_2=1$ as in the common practice)
\begin{equation}
\varepsilon=\frac{6e_{4}-2e_{2}}{6e_{4}+e_{2}}.
\label{eq:eps_moments}
\end{equation}
with $\varepsilon=0$ for $e_4=1/3$ and $\varepsilon\to1$ as $e_4\to\infty$. By making use of the mechanic equilibrium condition $P_\text{N}(x)=p_0$ we can write the square of the spatial derivative of the pseudopotential function
\begin{equation}
\left[\frac{\text{d}\psi}{\text{d}x}\right]^{2}=\frac{8\left(1-\varepsilon\right)}{Gc_{s}^{2}e_{2}}\psi^{\varepsilon}\int_{n_{g}}^{n}\text{d}\bar{n}\frac{\psi'}{\psi^{\varepsilon+1}}\left(p_{0}-nc_{s}^{2}-\frac{Gc_{s}^{2}e_{2}}{2}\psi^{2}\right).
\label{eq:PSIProfile}
\end{equation}
Given that in the bulk the first derivative is vanishing one ends up with an integral constraint
\begin{equation}
\int_{n_{g}}^{n_{l}}\text{d}\bar{n}\frac{\psi'}{\psi^{\varepsilon+1}}\left(p_{0}-nc_{s}^{2}-\frac{Gc_{s}^{2}e_{2}}{2}\psi^{2}\right)=0,
\label{eq:SCMaxwell}
\end{equation}
which can be used to compute the equilibrium pressure $p_0$ and densities $n_g$ and $n_l$ at a given \emph{temperature} $T=1/|G|$. Eq.~\eqref{eq:SCMaxwell} is reminiscent of Maxwell's construction although with a different integration measure, i.e., $\mbox{d}\bar{n}\; \psi'/\psi^{\varepsilon + 1}$ for SC versus $\mbox{d}\bar{n}/\bar{n}^2$ for the latter. This difference is related to the possibility of defining a consistent chemical potential $\mu$ for the SC model~\cite{Sbragaglia_2011} and it is known in the literature as the issue of thermodynamic consistency (see~\cite{Khajepor_2015,khajepor2016multipseudopotential} and references within). In~\cite{Sbragaglia_2011} it was demonstrated that a particular choice of $\psi$ with parameters depending on the underlying forcing stencil:
\begin{equation}
\psi\left(n\right)=\left(\frac{n}{\varepsilon+n}\right)^{\frac{1}{\varepsilon}},
\label{eq:psieps}
\end{equation}
the measure in Eq.~\eqref{eq:SCMaxwell} would always be the correct one.
Although this choice may seem stringent it was shown that in the multi-pseupotential approach~\cite{khajepor2016multipseudopotential} one can use combinations of the above pseudopotential with different parameters to recover widely used equations of state, e.g. Van der Waals or Carnahan-Starling. 

The surface tension is defined by
\begin{equation}
\sigma_{0}=\int_{-\infty}^{+\infty}\text{d}x\,\left[P_{\text{N}}\left(x\right)-P_{\text{T}}\left(x\right)\right].
\label{eq:sigma0pnpt}
\end{equation}
We consider the Taylor expansion of the difference $P_{\text{N}}\left(x\right)-P_{\text{T}}\left(x\right)$:
\begin{equation}
P_{\text{N}}\left(x\right)-P_{\text{T}}\left(x\right)\simeq Gc_{s}^{2}\left[c_{02}\psi\left(x\right)\frac{\text{d}^{2}\psi}{\text{d}x^{2}}+c_{11}\left(\frac{\text{d}\psi}{\text{d}x}\right)^{2}\right]
\label{eq:PNMTTaylor2nd}
\end{equation}
where the coefficients $c_{02}$ and $c_{11}$ are given by some linear combinations of the weights, i.e., some linear combinations of the isotropy coefficients. It is customary to truncate the expansion at second order as this allows one to recover results of the same functional form as in the Van der Waals theory~\cite{RowlinsonWidom82}. When performing the integral we can leverage the vanishing of the gradients in bulk, i.e., for $x\to\pm\infty$, and perform the integration by parts yielding
\begin{equation}
\sigma_{0}\simeq-\frac{Gc_{s}^{2}e_{4}}{2}\int_{-\infty}^{+\infty}\text{d}x\left(\frac{\text{d}\psi}{\text{d}x}\right)^{2},
\label{eq:scsigma}
\end{equation}
where we have used $c_{11}-c_{02}=-{e_{4}}/{2}$ for stencil of at least fourth order isotropy. All properties reviewed above have an analytical expression which can be numerically evaluated but this is not the case for the Tolman length for which a phenomenological approach has been used so far~\cite{Lulli_2022,lulli2022mesoscale,Hosseini_2022}, i.e., using simulations data for droplets (positive curvature) and bubbles (negative curvature) in order to evaluate the generalized surface tension~\cite{RowlinsonWidom82}
\begin{equation}
\sigma\left[R\right]=\int_{-\infty}^{+\infty}\text{d}r\,\left(\frac{r}{R}\right)^{d-1}\left[P_{\text{J}}\left(r;R\right)-P_{\text{T}}\left(r\right)\right]
\end{equation}
where we have defined $P_{\text{J}}\left(r;R\right)=P_{\text{in}}-\left(P_{\text{in}}-P_{\text{out}}\right)\theta\left(r-R\right)$. Here $\theta\left(x\right)$ the Heaviside function, $P_\text{in}$ and $P_\text{out}$ are the bulk pressure evaluated at the center of the droplet/bubble and far away, respectively and $P_\text{T}(r)=P^{yy}(x)=P^{zz}(x)$, with $\hat{r}=\hat{x}$. After interpolating the data for $\sigma[R]$ one can determine the position and the value of the minimum thus yielding the position of the surface of tension $R_s$ and the corresponding value of the surface tension $\sigma(R_s)$. The Tolman length has been estimated~\cite{Lulli_2022,lulli2022mesoscale,Hosseini_2022} by analyzing a sequence of bubbles/droplets of different sizes and combining together Eq.~\eqref{eq:ExactLaplace} and Eq.~\eqref{eq:SigmaCExpansion} in order to obtain at first order in $R_s^{-1}$
\begin{equation}
\frac{\Delta P\cdot R_{s}}{\left(d-1\right)\sigma_{0}}\simeq1-\frac{\left(d-1\right)\delta}{R_{s}},
\end{equation}
where $\Delta P = P_\text{in} - P_\text{out}$, i.e., $\delta$ can be estimated from the slope of the straight line obtained by fitting the data of $\Delta P\cdot R_s / (d-1) \sigma_0$ vs. $R_s^{-1}$ in the region $|R_s^{-1}|\ll1$. 

\section{Higher-order Tuning}\label{sec:met}
In this Section, we define our tuning approach which leverages higher-order derivatives allowing for an independent tuning of (i) the flat interface profile, (ii) surface tension and (iii) Tolman length. As it can be seen from the Taylor expansion of the SC force in Eq.~\eqref{eq:FSCTaylor} the moments of the forcing stencil play a crucial role. In the general case, it is always possible to split these moments into a fully-isotropic and an anisotropic contribution
\begin{equation}
\langle e^{\mu_{1}}\cdots e^{\mu_{2n}}\rangle_{W}=e_{2n}\Delta^{\mu_{1}\ldots\mu_{2n}}+\langle e^{\mu_{1}}\cdots e^{\mu_{2n}}\rangle_{W}^{\text{aniso}},
\end{equation}
where we refer to the set of coefficients $\{e_{2n}\}$ as \emph{isotropy coefficients}. Here, $\Delta^{\mu_{1}\ldots\mu_{2n}}$ is the rank-$2n$ fully isotropic tensor, i.e., the invariant tensor under the group of rotations, which can be recursively defined starting from $2n=2$ corresponding to the Kronecker delta $\delta^{\mu\nu}$; for $2n=4$ one has $\Delta^{\mu_1\mu_2\mu_3\mu_4}=\delta^{\mu_1\mu_2}\delta^{\mu_3\mu_4}+\delta^{\mu_1\mu_3}\delta^{\mu_4\mu_2}+\delta^{\mu_1\mu_4}\delta^{\mu_2\mu_3}$. For the general case one has~\cite{Wolfram1986}
\begin{equation}
\Delta^{\mu_1\mu_2\ldots\mu_{2n+2}}=\sum_{i_k\in\pi_{2,2n+2}}\delta^{\mu_1\mu_{i_1}}\Delta^{\mu_{i_2}\mu_{i_3}\ldots\mu_{i_{2n+1}}}
\end{equation}
where $\pi_{a,b} = \{(a, a + 1,\ldots,b-1,b), (a + 1, a + 2,\ldots,b, a),\ldots\}$ is the set of all the even permutation of the integers from $a$ to $b$. The anisotropic component can be written as a combination of non-isotropic tensors
\begin{equation}
\begin{split}\langle e^{\mu_{1}}\cdots e^{\mu_{2n}}\rangle_{W}^{\text{aniso}} & =I_{2n,0}\delta^{\mu_{1}\ldots\mu_{2n}}\\
+I_{2n,1} & \left[\delta^{\mu_{1}\mu_{2}}\delta^{\mu_{3}\ldots\mu_{2n}}+\text{perms.}\right]+\ldots
\end{split}
\end{equation}
where the symbol $\delta^{\mu_1\ldots\mu_{2k}}$ indicate the generalized rank-$2k$ Kronecker delta which is equal to one only if all the $2k$ indices are the same and vanishes otherwise, and where we refer to the set of coefficients $\{I_{2n,k}\}$ as the \emph{anisotropy coefficients}. The key property of the two sets $\{e_{2n}\}$ and $\{I_{2n,k}\}$ is that they can be expressed as linear combination of the stencil weights $\{W(|\mathbf{e}_a|^2)\}$. This was implicitly stated in~\cite{Wolfram1986}, however, without outlining a general strategy for the computation. The LBM literature~\cite{shan2006analysis,Sbragaglia_2007} mainly focused on determining maximally isotropic tensor for the SC forcing. Only recently, in~\cite{Lulli_2021}, a general approach for determining these linear combinations to an arbitrary order was provided for two-dimensional groups of vectors. In order to tackle the three-dimensional case we adopt the construction of the anisotropy coefficients outlined in equation (F2) of~\cite{Lulli_2021} in the spirit of~\cite{Sbragaglia_2007}, i.e., we calculate the sufficient conditions for the vanishing of the anisotropy coefficients $\{\bar{I}_{2n,k}\}$ at a given order and use them, together with the definition of the isotropy coefficients $\{e_{2n}\}$, in order to express the weights $W$ as linear combinations of $\{\bar{I}_{2n,k}\}$ and $\{e_{2n}\}$. Finally, we set $\{\bar{I}_{2n,k}\}$ to zero, i.e. we impose a given degree of isotropy (see Appendix~\ref{app:e4ex} for a practical example). The end result is that the weights are expressed only in terms of the forcing moments, i.e. $\{e_{2n}\}$. Given that the coefficients setting the values of the interface parameters can be expressed as linear combinations of the weights one can invert these relations and finally write $W$ directly as a function of the physical knobs, i.e., surface tension and Tolman length coefficients. We finally wish to remark that, once the two- or three-dimensional $W$'s are expressed in terms of $\{e_{2n}\}$, the linear combinations of the pressure tensor expansion coefficients do not change with the dimension for a fixed degree of isotopy (see Appendix~\ref{app:stencil}).

\subsection{Surface Tension}
Let us consider again the expansion of $P_\text{N}(x) - P_\text{T}(x)$ and truncate it now at fourth order
\begin{equation}
\begin{split}
&P_{\text{N}}\left(x\right)-P_{T}\left(x\right)\simeq Gc_{s}^{2}\left[c_{02}\psi\frac{\text{d}^{2}\psi}{\text{d}x^{2}}+c_{11}\left(\frac{\text{d}\psi}{\text{d}x}\right)^{2}\right]\\&\quad+Gc_{s}^{2}\left[c_{04}\psi\frac{\text{d}^{4}\psi}{\text{d}x^{4}}+c_{13}\frac{\text{d}\psi}{\text{d}x}\frac{\text{d}^{3}\psi}{dx^{3}}+c_{22}\left(\frac{\text{d}^{2}\psi}{\text{d}x^{2}}\right)^{2}\right]
\end{split}
\end{equation}
so that the surface tension integral reads
\begin{equation}
\begin{split}
&\sigma_{0}\simeq Gc_{s}^{2}\left(c_{11}-c_{02}\right)\int_{-\infty}^{+\infty}\text{d}x\left(\frac{\text{d}\psi}{\text{d}x}\right)^{2}\\&\qquad+Gc_{s}^{2}\left(c_{22}-c_{13}+c_{04}\right)\int_{-\infty}^{+\infty}\text{d}x\left(\frac{\text{d}^{2}\psi}{\text{d}x^{2}}\right)^{2}\\
&\qquad \simeq Gc_{s}^{2}\left(\hat{\sigma}_{0}\Sigma_{11}+\hat{\sigma}_{1}\Sigma_{22}\right)
\end{split}\label{eq:HOSigmaInt}
\end{equation}
where $\hat{\sigma}_{0}=c_{11}-c_{02}=-e_{4}/2$, $\hat{\sigma}_{1}=c_{22}-c_{13}+c_{04}$, $\Sigma_{kk}=\int_{-\infty}^{+\infty}\text{d}x\left(\frac{\text{d}^k\psi}{\text{d}x^k}\right)^{2}$. Given that the leading-order coefficient is $\hat{\sigma}_0 \propto e_4$ and that $\varepsilon = \varepsilon(e_4)$ controls the equilibrium flat-interface profile [see Eq.~\eqref{eq:PSIProfile}] it is not possible to tune the surface tension alone by changing the value of $e_4$. Hence, we wish to do so by changing the higher-order contribution proportional to $\Sigma_{22}$, i.e., by tuning the value of $\hat{\sigma}_1$. Another important question is whether or not we have full analytical control over the expression in Eq.~\eqref{eq:HOSigmaInt}, in particular over the second derivative of $\psi$ defining the integral $\Sigma_{22}$. In fact, the mechanic equilibrium condition and the analytic expression for interface profile derivative (see Eq.~\eqref{eq:SCMaxwell} and~\eqref{eq:PSIProfile}) are already enough to fully determine the second derivative as follows
\begin{equation}
\begin{split}
&\frac{\text{d}^{2}\psi}{\text{d}x^{2}}=\frac{1}{2}\frac{\text{d}}{\text{d}\psi}\left(\frac{\text{d}\psi}{\text{d}x}\right)^{2}\\&=\frac{1}{2\psi}\left\{ \varepsilon\left(\frac{\text{d}\psi}{\text{d}x}\right)^{2}+\frac{1}{Gc_{s}^{2}}\frac{8\left(1-\varepsilon\right)}{e_{2}}\left[p_{0}-nc_{s}^{2}-\frac{Gc_{s}^{2}e_{2}}{2}\psi^{2}\right]\right\}.
\end{split}\label{eq:d2psi}
\end{equation}
Hence, the higher-order approximation for $\sigma_0$ in Eq.~\eqref{eq:HOSigmaInt} is still fully under control from the analytical perspective within the usual approximation for the evaluation of the equilibrium properties of the flat-interface profile.

\subsection{Tolman Length}
We start from the one-dimensional definition of the Tolman length~\cite{Tolman1949,RowlinsonWidom82} as the difference between the positions of the equimolar surface $z_e$ and the surface of tension $z_s$, i.e.,
\begin{equation}
\delta = z_e - z_s,
\label{eq:zemzs}
\end{equation}
where $z_e$ is defined as the value where the \emph{adsorbance} vanishes, i.e., $\Gamma(z_e)=0$~\cite{RowlinsonWidom82}. A simple geometrical interpretation of the equimolar surface is that $z_e$ would be the position at which a sharp interface would rest at fixed total mass and bulk densities. In one dimension the position of the surface of tension can be defined using the pressure tensor as follows
\begin{equation}
\begin{split}z_{s}= & \frac{\int_{-\infty}^{+\infty}\text{d}x\,x\,\left[P_{\text{N}}\left(x\right)-P_{\text{T}}\left(x\right)\right]}{\int_{-\infty}^{+\infty}\text{d}x\,\left[P_{\text{N}}\left(x\right)-P_{\text{T}}\left(x\right)\right]}\\
 & =\frac{1}{\sigma_{0}}\int_{-\infty}^{+\infty}\text{d}x\,x\,\left[P_{\text{N}}\left(x\right)-P_{\text{T}}\left(x\right)\right],
\end{split}\label{eq:ZSDef}
\end{equation}
i.e., as the \emph{first moment} of the quantity $P_\text{N}(x) - P_\text{T}(x)$~\cite{RowlinsonWidom82}. We now write the fourth-order Taylor expansion for the integrand at the numerator of Eq.~\eqref{eq:ZSDef}, namely
\begin{equation}
\begin{split}
&x\left[P_{\text{N}}\left(x\right)-P_{T}\left(x\right)\right]\simeq Gc_{s}^{2}\left[c_{02}\,x\,\psi\frac{\text{d}^{2}\psi}{\text{d}x^{2}}+c_{11}\,x\,\left(\frac{\text{d}\psi}{\text{d}x}\right)^{2}\right]\\&\quad+Gc_{s}^{2}\left[c_{04}\,x\,\psi\frac{\text{d}^{4}\psi}{\text{d}x^{4}}+c_{13}\,x\,\frac{\text{d}\psi}{\text{d}x}\frac{\text{d}^{3}\psi}{dx^{3}}+c_{22}\,x\,\left(\frac{\text{d}^{2}\psi}{\text{d}x^{2}}\right)^{2}\right].
\end{split}
\end{equation}
Our aim is perform integration by parts until we obtain terms proportional to the square of the first and second derivatives. As boundary conditions we will be using the fact that for a flat interface all derivatives identically vanish in the bulk, i.e., $\mbox{d}^k \psi / \mbox{d} x^k |_\text{bulk} = 0$ for $k\geq 1$, hence we automatically discard all total derivatives containing these terms. The only terms we need to consider are those proportional to $c_{02}$, $c_{04}$ and $c_{13}$, yielding
\begin{equation}
\begin{split}
&c_{02}\,x\,\psi\frac{\text{d}^{2}\psi}{\text{d}x^{2}}=-c_{02}\left[\frac{1}{2}\frac{\text{d}}{\text{d}x}\psi^{2}+x\left(\frac{\text{d}\psi}{\text{d}x}\right)^{2}\right],\\
&c_{04}\,x\,\psi\frac{\text{d}^{4}\psi}{\text{d}x^{4}}=c_{04}\,x\,\left(\frac{\text{d}^{2}\psi}{\text{d}x^{2}}\right)^{2},\\
&c_{13}\,x\,\frac{\text{d}\psi}{\text{d}x}\frac{\text{d}^{3}\psi}{\text{d}x^{3}}=-c_{13}\,x\,\left(\frac{\text{d}^{2}\psi}{\text{d}x^{2}}\right)^{2}.
\end{split}
\end{equation}

Summing all the terms we obtain
\begin{equation}
\begin{split}
&z_{s}=-\frac{Gc_{s}^{2}c_{02}}{2\sigma_{0}}\int_{-\infty}^{+\infty}\text{d}x\,\frac{\text{d}}{\text{d}x}\psi^{2}\\&+\frac{Gc_{s}^{2}}{\sigma_{0}}\int_{-\infty}^{+\infty}\text{d}x\,\left[\left(c_{11}-c_{02}\right)\,x\,\left(\frac{\text{d}\psi}{\text{d}x}\right)^{2}\right]\\&+\frac{Gc_{s}^{2}}{\sigma_{0}}\int_{-\infty}^{+\infty}\text{d}x\,\left[\left(c_{22}-c_{13}+c_{04}\right)\,x\,\left(\frac{\text{d}^{2}\psi}{\text{d}x^{2}}\right)^{2}\right].
\end{split}
\end{equation}
The last two terms clearly match the structure of the expansion of the surface tension in Eq.~\eqref{eq:HOSigmaInt}, which is written in terms of integrals of positive quantities, hence, we can consider the last two terms as the first moment of a linear combination of two probability distributions, a monomodal one, given by $(\mbox{d}\psi / \mbox{d}x)^2$, and a bimodal one given by $(\mbox{d}^2\psi / \mbox{d}x^2)^2$ (the bimodality is due to the existence of an inflection point in the flat interface profile). Roughly speaking, such an average will fall at the middle point of the interface which in turn is a reasonable estimation of the position of the equimolar surface $z_e$. Hence, we can write for the estimation for the position of the surface of tension as
\begin{equation}
z_{s}\simeq z_{e}-\frac{Gc_{s}^{2}c_{02}}{2\sigma_{0}}\int_{-\infty}^{+\infty}\text{d}x\,\frac{\text{d}}{\text{d}x}\psi^{2},
\end{equation}
and after performing the integration of the total derivative we obtain the following estimation for the Tolman length for the one-dimensional interface
\begin{equation}
\delta=z_{e}-z_{s}\simeq-\frac{Gc_{s}^{2}}{2\sigma_{0}}\hat{\delta}_{0}\left[\psi^{2}\left(n_{l}\right)-\psi^{2}\left(n_{g}\right)\right],
\label{eq:SCTolmanFlat}
\end{equation}
with $\hat{\delta}_{0}=c_{02}$. This is the first analytical estimate of the Tolman length in the SC model. 

\subsection{Computing the Weights}
Now, we wish to show the difference of the present approach for determining the values of the weights with respect to the previous ones. Indeed, most literature focuses on defining forcing stencils where the degree of isotropy is \emph{maximal} for the given number of symmetry groups selected. In the seminal papers~\cite{shan2006analysis,Sbragaglia_2007} the degree of isotropy was connected to the magnitude of the spurious currents near the interface, being weaker for larger degree of isotropy. Several maximally isotropic stencils were provided in two and three dimensions. This amounts to writing the $N_\text{W}$ weights as a function of $e_2$ and $N_\text{W} - 1$ anisotropy coefficients and then set them to zero. In the case of the fourth-order isotropic forcing in two dimensions (see Appendix~\ref{app:e4ex}) one obtains $W(1)=1/3$ and $W(2)=1/12$~\cite{Shan_2008}. The above strategy can be extended to using more symmetry groups of vectors thus allowing for a higher isotropy order, however, once the symmetry groups are fixed, there is only one possible maximally isotropic stencil by construction (assuming all the anisotropy coefficients at each selected order can be set to zero). Furthermore, different maximally isotropic stencil lead to different values of the moments, e.g. $e_{4}$, hence different equilibrium densities, through $\varepsilon(e_4)$, and surface tension values (see Eq.~\eqref{eq:scsigma}), while keeping the equation of state unchanged. Hence, physical properties and maximal isotropy are inseparably intertwined (see Appendix~\ref{app:epse4} for a detailed discussion also with respect to~\cite{Li_2013,Khajepor_2015}). This is the fundamental reason why the approaches proposed in~\cite{Sbragaglia_2007,Falcucci_2007,Chibbaro_2008} eventually did not allow for a fully decoupled tuning of interface profile and surface tension~\cite{Li_2013,Khajepor_2015}. Indeed, it can be shown that our present approach can retrieve and improve upon the previously developed multi-range and multi-belt approaches as discussed in Appendix~\ref{app:MBMR}.

Indeed, one may choose to use the degrees of freedom encoded in the weights differently: rather than using $N_{\text{W}} - 1$ anisotropy coefficients one can resort to a smaller number while employing the definition of higher order isotropy coefficients $e_{2n}$, so that, the stencil is not maximally isotropic but can allow for the tuning of the selected $e_{2n}$'s. This approach was adopted in~\cite{Lulli_2021,lulli2024metastable} to tune the physical properties of the interface and the static structure factor independently on the degree of isotropy of the stencil. In this work we adapt the same strategy to demonstrate how the Tolman length can be tuned independently from the flat-interface profile and the surface tension with high precision: all we need to do is to compute the coefficients $\hat{\sigma}_0=c_{11} - c_{02}$, $\hat{\sigma}_1 = c_{22} - c_{13} + c_{04}$ and $\hat{\delta}_0 = c_{02}$ as a function of the isotropy coefficients, i.e., the moments, invert the expressions and substitute in the solution of the weights. After this procedure, the stencil weight will depend on the \emph{knobs} $\hat{\sigma}_0$, $\hat{\sigma}_1$ and $\hat{\delta}_0$.

\section{Results}\label{sec:results}
\subsection{Leading order surface tension tuning - $\hat{\sigma}_0$}\label{ssec:s0}
\begin{figure}[!t]
\includegraphics[scale=0.54]{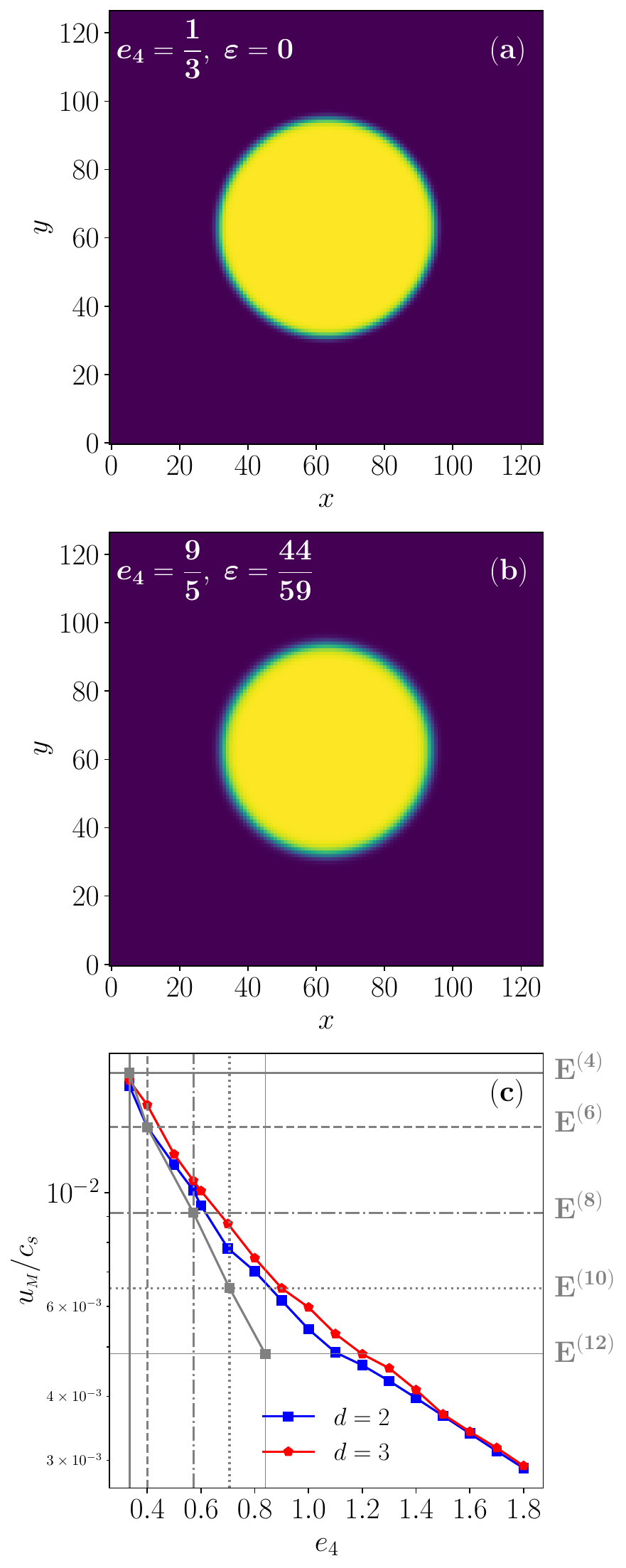}
\caption{Panel $(a)$ and $(b)$: density map for a two-dimensional droplet obtained using the weights reported in Table~\ref{tab:e4_tuning_6} for different values of $e_4$ and $\varepsilon$ yielding a thinner and thicker interface respectively. Panel $(c)$: largest Mach number for spurious currents as a function of $e_4$ with the values from higher isotropy stencils in two dimensions.}\label{fig:u_max_e4}
\end{figure}
\renewcommand{\arraystretch}{1.3}
\begin{table}[t!]
  \begin{ruledtabular}
  \centering
  \begin{tabular}{lccccc}
  & $E^{(4)}$ & $E^{(6)}$ & $E^{(8)}$ & $E^{(10)}$ & $E^{(12)}$\\
  \hline
  $e_4$ & 1/3 & 2/5 & 4/7 & 12/17 & 120/143 \\
  $e_6$ & 1/9 & 2/15 & 32/105 & 8/17 & 96/143 \\
  $e_8$ & 1/27 & 2/45 & 16/105 & 176/595 & 512/1001 \\
  $e_{10}$ & 1/135 & 2/225 & 16/225 & 104/595 & 1856/5005 \\
  $e_{12}$ & 1/675 & 2/1125 & 16/525 & 288/2975 & 256/1001 \\
  $N_e,\;d=2$ & 8 & 12 & 24 & 36 & 56 \\
  $N_e,\;d=3$ & 18 & 32 & 92 & 170 & 284 \\
  \end{tabular}
  \end{ruledtabular}
  \caption{Values for the isotropy constants of the two- and three-dimensional fully isotropic forcing stencils up to 12$^{th}$ order~\cite{Sbragaglia_2007}. The rows with $N_e$ indicate the number of vectors in the stencil.}\label{tab:e2n_values}
\end{table}  

\renewcommand{\arraystretch}{2.1}
\begin{table}[t!]
  \begin{ruledtabular}
  \centering
  \begin{tabular}{lcc}
  \multicolumn{3}{c}{Tuning $\mathbf{e_4} = -2\boldsymbol{\hat{\sigma}_0}$ -- 6$^{th}$-order Forcing Isotropy} \\
  \hline
  & \multicolumn{1}{c}{$\displaystyle d=2$} & \multicolumn{1}{c}{$\displaystyle d=3$} \\
  \hline
  $\displaystyle W(1)$
  & $\displaystyle \frac{8}{15}+\frac{4\hat{\sigma}_{0}}{3}$ 
  & $\displaystyle \frac{4}{15}+\frac{2\hat{\sigma}_{0}}{3}$ \\
  $\displaystyle W(2)$ 
  & $\displaystyle -\frac{\hat{\sigma}_{0}}{6}+\frac{1}{15}$ 
  & $\displaystyle \frac{2}{15}+\frac{\hat{\sigma}_{0}}{3}$ \\
  $\displaystyle W(3)$ & -- 
  & $\displaystyle -\frac{\hat{\sigma}_{0}}{4}-\frac{1}{30}$ \\
  $\displaystyle W(4)$ 
  & $\displaystyle -\frac{5\hat{\sigma}_{0}}{24}-\frac{1}{30}$ 
  & $\displaystyle -\frac{\hat{\sigma}_{0}}{6}-\frac{1}{40}$ \\
  $\displaystyle W(8)$ 
  & $\displaystyle -\frac{\hat{\sigma}_{0}}{48}-\frac{1}{240}$ 
  & $\displaystyle -\frac{\hat{\sigma}_{0}}{48}-\frac{1}{240}$\\
  $N_e$ & 16 & 44 \\
  \end{tabular}
  \end{ruledtabular}
  \caption{Weights for the 6$^{th}$ order isotropic forcing stencils in two and three dimensions as a function of the coefficient $\hat{\sigma}_0 = -e_4/2$. The three dimensional weights exactly project onto the two-dimensional ones. When selecting the value $e_4 = 2/5$ one obtains the usual $E^{(6)}$ weights (see Table~\ref{tab:e2n_values}). The rows with $N_e$ indicate the number of vectors in the stencil.}\label{tab:e4_tuning_6}
\end{table}

In this subsection we present the results for the leading-order tuning of the surface tension and its relation to the intensity of the spurious currents. In order to tune $\hat{\sigma}_0$ we need to tune $e_4$. This procedure does not ensure that the flat interface profile is unchanged. However, we will use these results in order to define reference stencils for the higher-order tuning of the surface tension through $\hat{\sigma}_1$ and of the Tolman length for $\hat{\delta}_0$. In order to tune $\hat{\sigma}_0$ while ensuring a 6$^{th}$ order isotropic forcing we need to use two linear combinations for $e_2$ and $e_4$ and two or three anisotropy coefficients in two or three dimensions (see Appendix~\ref{app:stencil} for the definition of the coefficients). Hence, we need a total of 4 weights in two dimensions, while 5 weights in dimension three. We do this by choosing the smallest possible symmetry groups, i.e., those with the smallest $\ell^2$ such that the equations are linearly independent. The resulting expressions are reported in Table~\ref{tab:e4_tuning_6}, after having set the anisotropy coefficients $\bar{I}_{4,0},\bar{I}_{6,0}$ and $\bar{I}_{6,1}$ (for $d=3$) to zero and $e_2=1$. We also report the total number of vectors in the stencil as $N_e$ for each dimension.

In Fig.~\ref{fig:u_max_e4}$(a)$ and $(b)$ we report the density field of two-dimensional droplets for two extreme values $e_4=1/3$ and $e_4=9/5$: as it can be seen the interface in panel $(a)$ is sharper than the one in $(b)$, because of the smaller value of $\varepsilon$ yielding a larger value of the first derivative, i.e., a steeper flat-interface profile (see Eq.~\eqref{eq:PSIProfile}). At the same time, it is well known that the interface width is related to the intensity of the spurious currents~\cite{Sbragaglia_2007,khajepor2016multipseudopotential}, and indeed in Fig.~\ref{fig:u_max_e4}$(c)$ we observe a roughly exponential reduction of the maximum velocity value $u_\text{M}$ as for increasing $e_4$, i.e., for thicker interfaces. It is interesting to compare the results for the two stencils defined in Table~\ref{tab:e4_tuning_6} with those obtained by the maximally isotropic forcing~\cite{Sbragaglia_2007}: indeed, the maximally isotropic cases show a faster decay of the spurious currents coming at the cost of a rapidly increasing number of vectors as reported in Table~\ref{tab:e2n_values}, while the decrease obtained as a function of $e_4$ alone is obtained using a constant number of vectors, $N_e = 16$ and $N_e = 44$ for two and three dimensions, respectively. These numbers are substantially smaller than the 12$^{th}$ order isotropic stencils $E^{(12)}$ with $N_e = 56$ and $N_e=284$ vectors in two and three dimensions, respectively, i.e., a factor of $3.5$ and $6.4$ less memory reads in comparison for the $e_4$-tuned stencils. We remark that the reduction obtained by tuning $e_4$ is roughly of one order of magnitude. Indeed, there are seemingly two different mechanisms at play for the reduction of the spurious currents, namely (i) a higher degree of isotropy and (ii) a higher value of $e_4$. Starting with a higher isotropic forcing stencil, i.e., with more vector symmetry groups used to implement higher-order isotropy conditions, would yield even better results, always at the cost of a larger stencil. The analysis we just performed will be instrumental for the results in the remainder of the paper.

\subsection{Higher-order surface tension tuning - $\hat{\sigma}_1$}\label{ssec:s1}
\renewcommand{\arraystretch}{2.1}
\begin{table}[t!]
  \begin{ruledtabular}
  \centering	
  \begin{tabular}{lcc}
  \multicolumn{3}{c}{Tuning $\mathbf{e_4} = -2\boldsymbol{\hat{\sigma}_0}$, $\mathbf{e_6} = 4\boldsymbol{\hat{\sigma}_1}$ -- 8$^{th}$-order Forcing Isotropy} \\
  \hline
  & \multicolumn{1}{c}{$\displaystyle d=2$} & \multicolumn{1}{c}{$\displaystyle d=3$} \\
  \hline
  $\displaystyle W(1)$
  & $\displaystyle \frac{59\hat{\sigma}_{0}}{20}+4\hat{\sigma}_{1}+\frac{51}{70}$ 
  & $\displaystyle \frac{1789 \hat{\sigma}_0}{420} + \frac{121 \hat{\sigma}_1}{18} + \frac{4}{5}$ \\
  
  $\displaystyle W(2)$ 
  & $\displaystyle -\frac{187\hat{\sigma}_{0}}{240}-\frac{41\hat{\sigma}_{1}}{24}-\frac{1}{280}$ 
  & $\displaystyle -\frac{11\hat{\sigma}_{0}}{28}-\frac{65\hat{\sigma}_{1}}{72}$ \\
  
  $\displaystyle W(3)$ 
  & --  
  & $\displaystyle -\frac{\hat{\sigma}_{0}}{5}-\frac{7\hat{\sigma}_{1}}{15}$ \\
  
  $\displaystyle W(4)$ 
  & $\displaystyle -\frac{5\hat{\sigma}_{0}}{8}-\hat{\sigma}_{1}-\frac{3}{35}$ 
  & $\displaystyle -\frac{1181\hat{\sigma}_{0}}{1680}-\frac{109\hat{\sigma}_{1}}{90}-\frac{1}{10}$ \\
  
  $\displaystyle W(5)$
  & $\displaystyle \frac{7\hat{\sigma}_{0}}{60}+\frac{\hat{\sigma}_{1}}{3}+\frac{1}{70}$ 
  & $\displaystyle -\frac{\hat{\sigma}_{0}}{56}-\frac{\hat{\sigma}_{1}}{72}$ \\
  
  $\displaystyle W(6)$
  & -- 
  & $\displaystyle \frac{2\hat{\sigma}_{0}}{105}+\frac{31\hat{\sigma}_{1}}{360}$ \\

  $\displaystyle W(8)$ 
  & $\displaystyle -\frac{13\hat{\sigma}_{0}}{240}-\frac{\hat{\sigma}_{1}}{12}-\frac{1}{112}$ 
  & $\displaystyle \frac{9\hat{\sigma}_{0}}{1120}+\frac{11\hat{\sigma}_{1}}{360}$\\

  $\displaystyle W_{(2,2,1)}(9)$
  & -- 
  & $\displaystyle -\frac{\hat{\sigma}_{0}}{1680}-\frac{\hat{\sigma}_{1}}{720}$ \\

  $\displaystyle W(9)$ 
  & $\displaystyle \frac{59\hat{\sigma}_{0}}{1620}+\frac{2\hat{\sigma}_{1}}{27}+\frac{1}{210}$ 
  & $\displaystyle \frac{103\hat{\sigma}_{0}}{1134}+\frac{281\hat{\sigma}_{1}}{1620}+\frac{4}{315}$\\

  $\displaystyle W(18)$ 
  & $\displaystyle \frac{17\hat{\sigma}_{0}}{6480}+\frac{\hat{\sigma}_{1}}{216}+\frac{1}{2520}$ 
  & $\displaystyle -\frac{\hat{\sigma}_{0}}{11340}-\frac{\hat{\sigma}_{1}}{3240}$\\

  $\displaystyle W(16)$ 
  & -- 
  & $\displaystyle -\frac{41\hat{\sigma}_{0}}{6720}-\frac{\hat{\sigma}_{1}}{90}-\frac{1}{1120}$\\
  $N_e$ & 32 & 140 \\
  \end{tabular}
  \end{ruledtabular}
  \caption{Weights for the 8$^{th}$-order isotropic forcing with tunable $\hat{\sigma}_0$ and $\hat{\sigma}_1$ in two and three dimensions, where $N_e$ is the total number of stencil vectors.}\label{tab:e4e6_tuning_8}
\end{table}
\begin{figure}[!t]
\includegraphics[scale=0.3825]{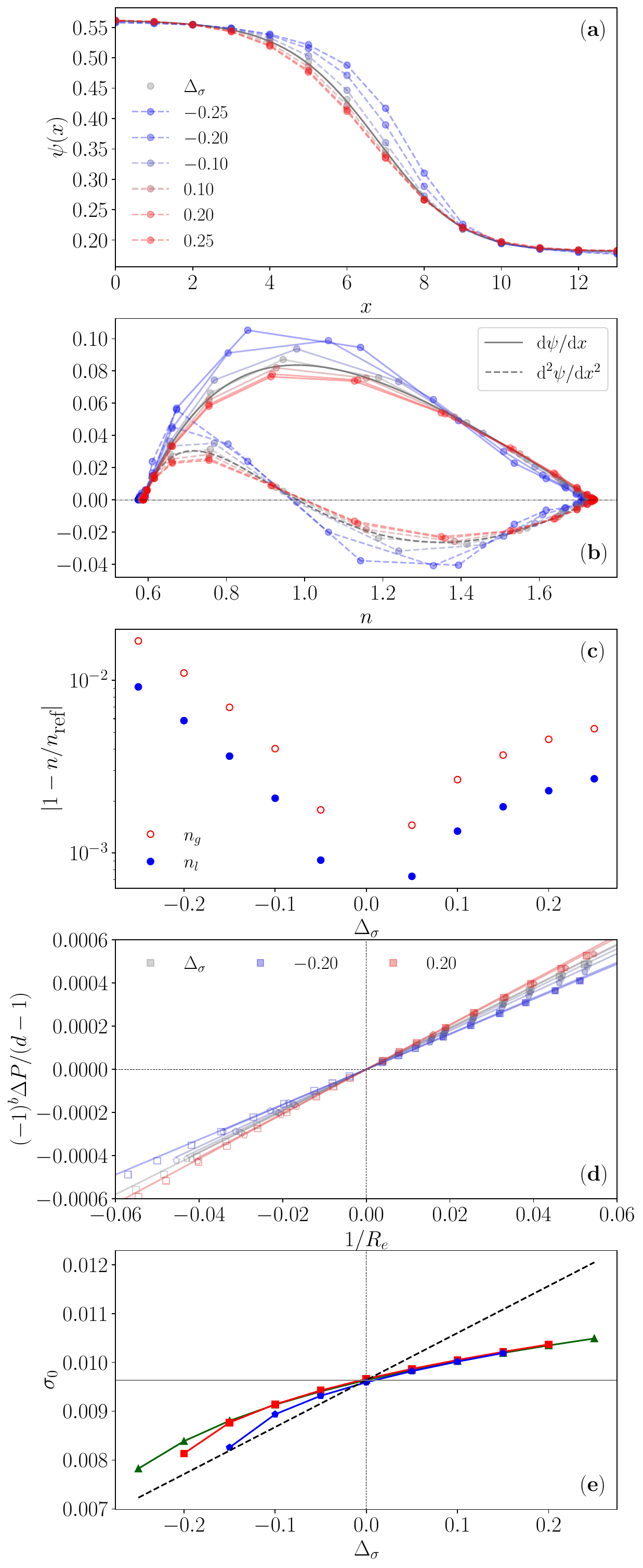}
\caption{Results for $\Delta\hat{T}=0.1$. Panels $(a)$, $(b)$ and $(c)$: one-dimensional simulation results for interface profiles, derivatives and equilibrium density changes as a function of $\Delta_\sigma$, respectively. Panels $(d)$ and $(e)$: two- and three-dimensional simulations results for the Laplace law and comparison of the estimated surface tension (triangles $d=1$, squares $d=2$ and pentagons $d=3$) with respect to the theoretical expectation in dashed.}\label{fig:sigma0p1}
\end{figure}
\begin{figure}[!t]
\includegraphics[scale=0.3825]{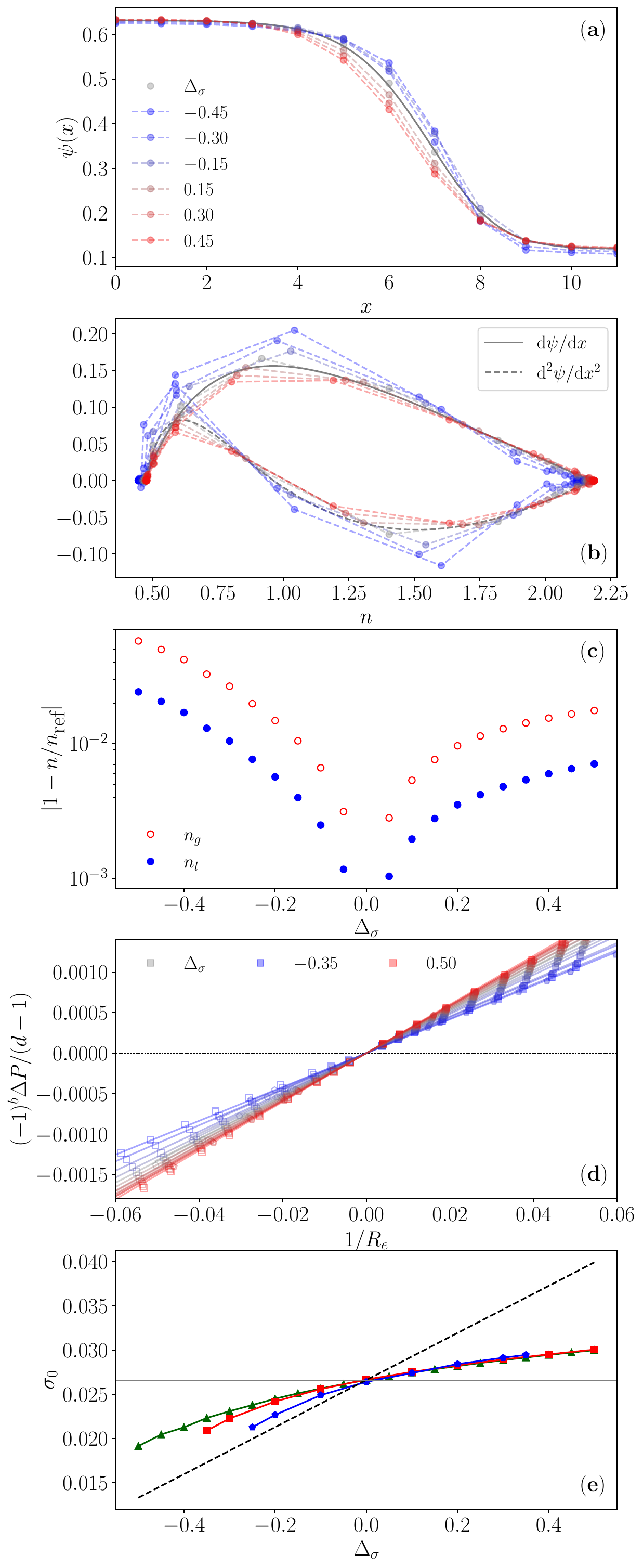}
\caption{Results for $\Delta\hat{T}=0.2$. Panels $(a)$, $(b)$ and $(c)$: one-dimensional simulation results for interface profiles, derivatives and equilibrium density changes as a function of $\Delta_\sigma$, respectively. Panels $(d)$ and $(e)$: two- and three-dimensional simulations results for the Laplace law and comparison of the estimated surface tension (triangles $d=1$, squares $d=2$ and pentagons $d=3$) with respect to the theoretical expectation in dashed.}\label{fig:sigma0p2}
\end{figure}
In this subsection we discuss the results related to the higher-order tuning of the surface tension by means of $\hat{\sigma}_1$ rather than $\hat{\sigma}_0$, i.e., by tuning the prefactor of $\Sigma_{22}$, i.e., the integral of the square of the second derivative. This approach does not rely on the tuning of $e_4$, and hence $\varepsilon$, so that the flat-interface equilibrium densities would be unchanged. In order to do this we need to resort to more vector symmetry groups than what we used in the previous subsection. Indeed, in order to maintain the same symmetry groups necessary for the tuning of the Tolman length, we select the groups $\ell^2\in\{1,2,4,5,8,9,18\}$ in two dimensions and the groups $\ell^2 \in \{1,2,3,4,5,6,8,9,16,18\}$ in three dimensions (with $\ell^2=9$ labeling the two groups related to the vectors $(2,2,1)$ and $(3,0,0)$). The number of groups necessary in three dimensions is larger because the number of anisotropy coefficients at a given order is larger with respect to the two-dimensional case. With this choice, in order to obtain independent combinations for $\hat{\sigma}_0$ and $\hat{\sigma}_1$, we need at least 6$^{th}$ order forcing isotropy and to include the isotropy constant $e_6$ in the definition of the weights. With these requirements the two coefficients can be written as
\begin{equation}
\hat{\sigma}_{0}=-\frac{e_{4}}{2},\qquad\hat{\sigma}_{1}=\frac{e_{6}}{4},
\label{eq:hatsigma01eI}
\end{equation}
which are clearly independent. Given that we are using three isotropy coefficients to solve the weights, namely $e_2$, $e_4$ and $e_6$, we can use $N_{\text{W}} - 3 = 4$ anisotropy coefficients in two dimensions and $N_\text{W} - 3=8$ in three dimensions, allowing for the imposition of $8^{th}$ order isotropy for both dimensions. We report in Table~\ref{tab:e4e6_tuning_8} the expressions for the weights as a function of $\hat{\sigma}_0$ and $\hat{\sigma}_1$. After setting $\hat{\sigma}_0=-2e_4=-2/3$, as in the single-belt $E^{(4)}$ case, we aim at tuning the surface tension only through $\hat{\sigma}_1$ whose reference value is set to satisfy a dimensional scaling, i.e., $e_6=e_4^{3/2}$. In this case, $\varepsilon=0$ so that the thermodynamic consistent pseudo-potential function is given by $\psi=\exp{(-1/n)}$ (see Section~\ref{ssec:interface} and~\cite{Sbragaglia_2011}). It is useful to define a tuning parameter $\Delta_\sigma$ with respect to the reference value $\sigma_{0}\simeq Gc_{s}^{2}\left(\hat{\sigma}_{0}\Sigma_{11}+\hat{\sigma}_{1}\Sigma_{22}\right)$ (see Eq.~\eqref{eq:HOSigmaInt}) of the surface tension at a given temperature. Given that we are using a higher-order tuning we need to consider the full expression for $\sigma_0$ after the change $\hat{\sigma}_1 \to \hat{\sigma}_1 + \Delta\hat{\sigma}_1$
\begin{equation}
\begin{split}
&\sigma_{0}\left(\Delta_{\sigma}\right)\simeq Gc_{s}^{2}\left(\hat{\sigma}_{0}\Sigma_{11}+\hat{\sigma}_{1}\Sigma_{22}+\Delta\hat{\sigma}_{1}\Sigma_{22}\right)\\&=\sigma_{0}\left(1+\frac{\Sigma_{22}}{\hat{\sigma}_{0}\Sigma_{11}+\hat{\sigma}_{1}\Sigma_{22}}\Delta\hat{\sigma}_{1}\right)=\sigma_{0}\left(1+\Delta_{\sigma}\right),
\end{split}
\end{equation}
where we have defined $\Delta_{\sigma}$ as
\begin{equation}
\Delta_{\sigma}=\frac{\Delta\hat{\sigma}_{1}\Sigma_{22}}{\hat{\sigma}_{0}\Sigma_{11}+\hat{\sigma}_{1}\Sigma_{22}}.
\end{equation}
All results will be parametrized as a function of $\Delta_\sigma$. As shown in Equations~\eqref{eq:PSIProfile} and~\eqref{eq:d2psi}, it is possible to find analytical expressions for the first and second derivatives of the flat-interface profiles, so that one can numerically evaluate the integrals $\Sigma_{11}$ and $\Sigma_{22}$ and thus set the tuning scale. We run simulations at two temperatures, or equivalently, two values of $G$, such that $\Delta\hat{T} = G/G_c - 1 = 0.1, 0.2$, with the critical value $G_c = -e^2$ for $\psi=\exp{(-1/n)}$ and lower temperatures corresponding to larger values of $\Delta\hat{T}$.

We report in Fig.~\ref{fig:sigma0p1} and Fig.~\ref{fig:sigma0p2} the results of simulations for $d=1$ in panels $(a)$, $(b)$ and $(c)$, and $d=2,3$ in panels $(d)$ and $(e)$, for $\Delta\hat{T}=0.1$ and $\Delta\hat{T}=0.2$, respectively. Reference values are reported in gray. Comparing panels $(a)$ one can see that while at the interface some noticeable differences come about, as one can also see from the derivatives in panels $(b)$, in bulk the matching is quite accurate as confirmed by the results in panels $(c)$ showing variations with respect to the reference densities of the order of $10^{-2}$. Furthermore, the convergence of the derivatives appears more precise when considering $\Delta_\sigma > 0$ while it worsens for $\Delta_\sigma < 0$ for both temperatures. Indeed, the present approach, based on tuning higher-order terms of the surface tension expansion, seems to be able to keep the equilibrium density values unchanged to a good precision, differently from what reported in~\cite{Li_2013} in the case of multi-range/belt tuning. In panels $(d)$ we show the results for the Laplace law, using the equimolar radius $R_e$, as defined by the geometric construction
\begin{equation}
\begin{split}\pi R_{e}^{2}n_{l}+\left(L^{2}-\pi R_{e}^{2}\right)n_{g}=M, & \;\text{for}\;d=2,\\
\frac{4}{3}\pi R_{e}^{3}n_{l}+\left(L^{3}-\frac{4}{3}\pi R_{e}^{3}\right)n_{g}=M, & \;\text{for}\;d=3,
\end{split}
\end{equation}
where $M$ is the total mass of the system, which is analogous to the definition of $z_e$ as the point of vanishing adsorbance $\Gamma(z_e)=0$ in one dimension. We simulated a sequence of square/cubic systems of linear size $L$ keeping the initial droplet/bubble size at a constant ratio $L/R=4$ and evaluated $\Delta P=P_\text{in} - P_\text{out}$ where $P_\text{in}$ is the bulk pressure at the center of mass while $P_\text{out}$ is the bulk pressure at one of the system corners. In panels $(d)$ we report the data for both droplets (positive curvature) and bubbles (negative curvature) as it is custom for the Tolman length analysis, however, in this case we need to introduce a sign change in the pressure for the bubbles which is obtained through the factor $(-1)^b$ where $b=1$ for bubble and $b=0$ for droplets. The results of the linear fits performed around $R_e^{-1}\simeq 0$ are reported in panels $(e)$, for both $d=2$ (squares) and $d=3$ (pentagons), along with the evaluation of the surface tension from the one-dimensional simulations (triangles) according to Eq.~\eqref{eq:sigma0pnpt} by means of the interpolation of the lattice pressure tensor components. Simulations results are matching for different dimensions in the region $\Delta_\sigma>0$ while they differ for $\Delta_\sigma < 0$. Furthermore, for both temperatures the estimated variation of the surface tension does not match the expected slope reported in dashed black, i.e., the variation is not just proportional to $\Delta_\sigma$.

In summary, the results of this subsection show that it is indeed possible to tune the surface tension only through an appropriate choice of the weights with a very limited effect on the equilibrium densities of the two phases. The tuning range increases as the temperature lowers, with results matching across dimensions in the region $\Delta_\sigma > 0$. The range of variation can be extended when modifying the forcing by inserting $F^2$ terms in the rank-2 tensor $T$ (see Eq.~\eqref{eq:GuoForcing}) as proposed in~\cite{Li_2013,khajepor2016multipseudopotential}, which would in turn allow for the values of $e_4$ and $\varepsilon$ to be changed independently (see Appendix~\ref{app:epse4} for more details).

\subsection{Higher-order Tolman length tuning - $\hat{\delta}_0$}\label{ssec:d0}
\renewcommand{\arraystretch}{2.1}
\begin{table}[t!]
  \begin{ruledtabular}
  \centering
  \begin{tabular}{lc}
  \multicolumn{2}{c}{Tuning $\boldsymbol{\hat{\sigma}_0}$, $\boldsymbol{\hat{\sigma}_1}$ and $\boldsymbol{\hat{\delta}_0}$ -- 6$^{th}$-order Forcing Isotropy} \\
  \hline
  & $\displaystyle d=2$ \\
  \hline
  $\displaystyle W(1)$
  & $\displaystyle \frac{9\hat{\delta}_{0}}{5}+\frac{47\hat{\sigma}_{0}}{12}+\frac{19\hat{\sigma}_{1}}{5}+\frac{3}{5}$ \\
  
  $\displaystyle W(2)$ 
  & $\displaystyle \frac{3\hat{\delta}_{0}}{10}-\frac{89\hat{\sigma}_{0}}{144}-\frac{209\hat{\sigma}_{1}}{120}-\frac{1}{40}$ \\
    
  $\displaystyle W(4)$ 
  & $\displaystyle \frac{9\hat{\delta}_{0}}{10}-\frac{17\hat{\sigma}_{0}}{120}-\frac{11\hat{\sigma}_{1}}{10}-\frac{3}{20}$ \\
  
  $\displaystyle W(5)$
  & $\displaystyle -\frac{6\hat{\delta}_{0}}{5}-\frac{19\hat{\sigma}_{0}}{36}+\frac{7\hat{\sigma}_{1}}{15}+\frac{1}{10}$ \\
  
  $\displaystyle W(8)$ 
  & $\displaystyle \frac{3\hat{\delta}_{0}}{4}+\frac{251\hat{\sigma}_{0}}{720}-\frac{\hat{\sigma}_{1}}{6}-\frac{1}{16}$ \\

  $\displaystyle W(9)$ 
  & $\displaystyle \frac{\hat{\delta}_{0}}{15}+\frac{13\hat{\sigma}_{0}}{180}+\frac{\hat{\sigma}_{1}}{15}$ \\

  $\displaystyle W(18)$ 
  & $\displaystyle -\frac{\hat{\delta}_{0}}{30}-\frac{11\hat{\sigma}_{0}}{720}+\frac{\hat{\sigma}_{1}}{120}+\frac{1}{360}$ \\
  \end{tabular}
  \end{ruledtabular}
  \caption{Weights for the 6$^{th}$-order isotropic forcing with tunable $\hat{\sigma}_0$, $\hat{\sigma}_1$ and $\hat{\delta}_0$ in two dimensions.}\label{tab:e4e6e8_tuning_6_2d}
\end{table}
\renewcommand{\arraystretch}{2.1}
\begin{table}[t!]
  \begin{ruledtabular}
  \centering
  \begin{tabular}{lc}
  \multicolumn{2}{c}{Tuning $\boldsymbol{\hat{\sigma}_0}$, $\boldsymbol{\hat{\sigma}_1}$ and $\boldsymbol{\hat{\delta}_0}$ -- 6$^{th}$-order Forcing Isotropy} \\
  \hline
  & $\displaystyle d=3$ \\
  \hline
  $\displaystyle W(1)$
  & $\displaystyle \frac{339\hat{\delta}_{0}}{10}+\frac{14359\hat{\sigma}_{0}}{720}-\frac{149\hat{\sigma}_{1}}{120}-\frac{81}{40}$ \\
  
  $\displaystyle W(2)$ 
  & $\displaystyle -\frac{27\hat{\delta}_{0}}{2}-\frac{107\hat{\sigma}_{0}}{16}+\frac{151\hat{\sigma}_{1}}{72}+\frac{9}{8}$ \\
  
  $\displaystyle W(3)$ 
  & $\displaystyle \frac{36\hat{\delta}_{0}}{5}+\frac{19\hat{\sigma}_{0}}{6}-\frac{61\hat{\sigma}_{1}}{30}-\frac{3}{5}$ \\
  
  $\displaystyle W(4)$ 
  & $\displaystyle -\frac{21\hat{\delta}_{0}}{10}-\frac{613\hat{\sigma}_{0}}{360}-\frac{49\hat{\sigma}_{1}}{60}+\frac{3}{40}$ \\
  
  $\displaystyle W(5)$
  & $\displaystyle -\frac{9\hat{\delta}_{0}}{20}-\frac{19\hat{\sigma}_{0}}{96}+\frac{137\hat{\sigma}_{1}}{720}+\frac{3}{80}$ \\
  
  $\displaystyle W(6)$
  & $\displaystyle -\frac{3\hat{\delta}_{0}}{10}-\frac{19\hat{\sigma}_{0}}{144}+\frac{41\hat{\sigma}_{1}}{360}+\frac{1}{40}$ \\

  $\displaystyle W(8)$ 
  & $\displaystyle \frac{9\hat{\delta}_{0}}{10}+\frac{199\hat{\sigma}_{0}}{480}-\frac{31\hat{\sigma}_{1}}{144}-\frac{3}{40}$\\

  $\displaystyle W_{(2,2,1)}(9)$
  & $\displaystyle -\frac{3\hat{\delta}_{0}}{40}-\frac{19\hat{\sigma}_{0}}{576}+\frac{7\hat{\sigma}_{1}}{288}+\frac{1}{160}$ \\

  $\displaystyle W(9)$ 
  & $\displaystyle \frac{11\hat{\delta}_{0}}{15}+\frac{17\hat{\sigma}_{0}}{40}-\frac{\hat{\sigma}_{1}}{60}-\frac{61}{1260}$\\

  $\displaystyle W(18)$ 
  & $\displaystyle -\frac{\hat{\delta}_{0}}{30}-\frac{11\hat{\sigma}_{0}}{720}+\frac{\hat{\sigma}_{1}}{120}+\frac{1}{360}$\\

  $\displaystyle W(16)$ 
  & $\displaystyle -\frac{3\hat{\delta}_{0}}{40}-\frac{29\hat{\sigma}_{0}}{720}+\frac{\hat{\sigma}_{1}}{120}+\frac{3}{560}$\\
  \end{tabular}
  \end{ruledtabular}
  \caption{Weights for the 6$^{th}$-order isotropic forcing with tunable $\hat{\sigma}_0$, $\hat{\sigma}_1$ and $\hat{\delta}_0$ in three dimensions.}\label{tab:e4e6e8_tuning_6_3d}
\end{table}
\begin{figure}[!t]
\includegraphics[scale=0.46]{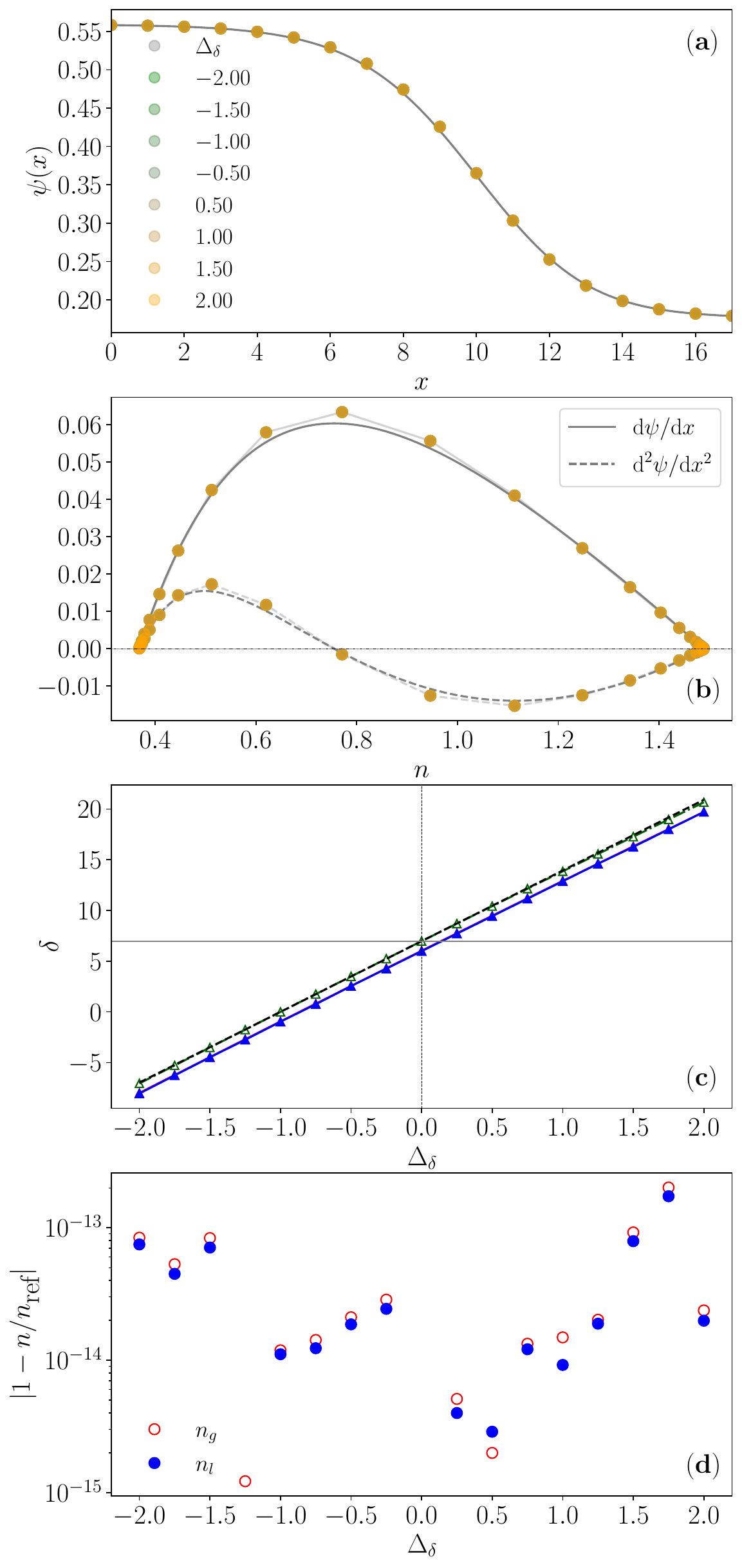}
\caption{Results for $\Delta\hat{T} = 0.1$ for flat interface simulations. Panel $(a)$: interface profile for several values of $\Delta_\delta$. Panel $(b)$: derivative profiles for the same values of $\Delta_\delta$ as in panel $(a)$. Panel $(c)$: different estimates of the Tolman length, according to Eq.~\eqref{eq:zemzs} and Eq.~\eqref{eq:deltaads} in red and blue triangles, and according to Eq.~\eqref{eq:SCTolmanFlat} in empty green triangles. The expected tuning behavior is reported in dashed black. Panel $(d)$: relative variation of the equilibrium bulk densities. Data in gray correspond to the reference value $\Delta_\delta=0$.}\label{fig:flat0p1}
\end{figure}
\begin{figure}[!t]
\includegraphics[scale=0.44]{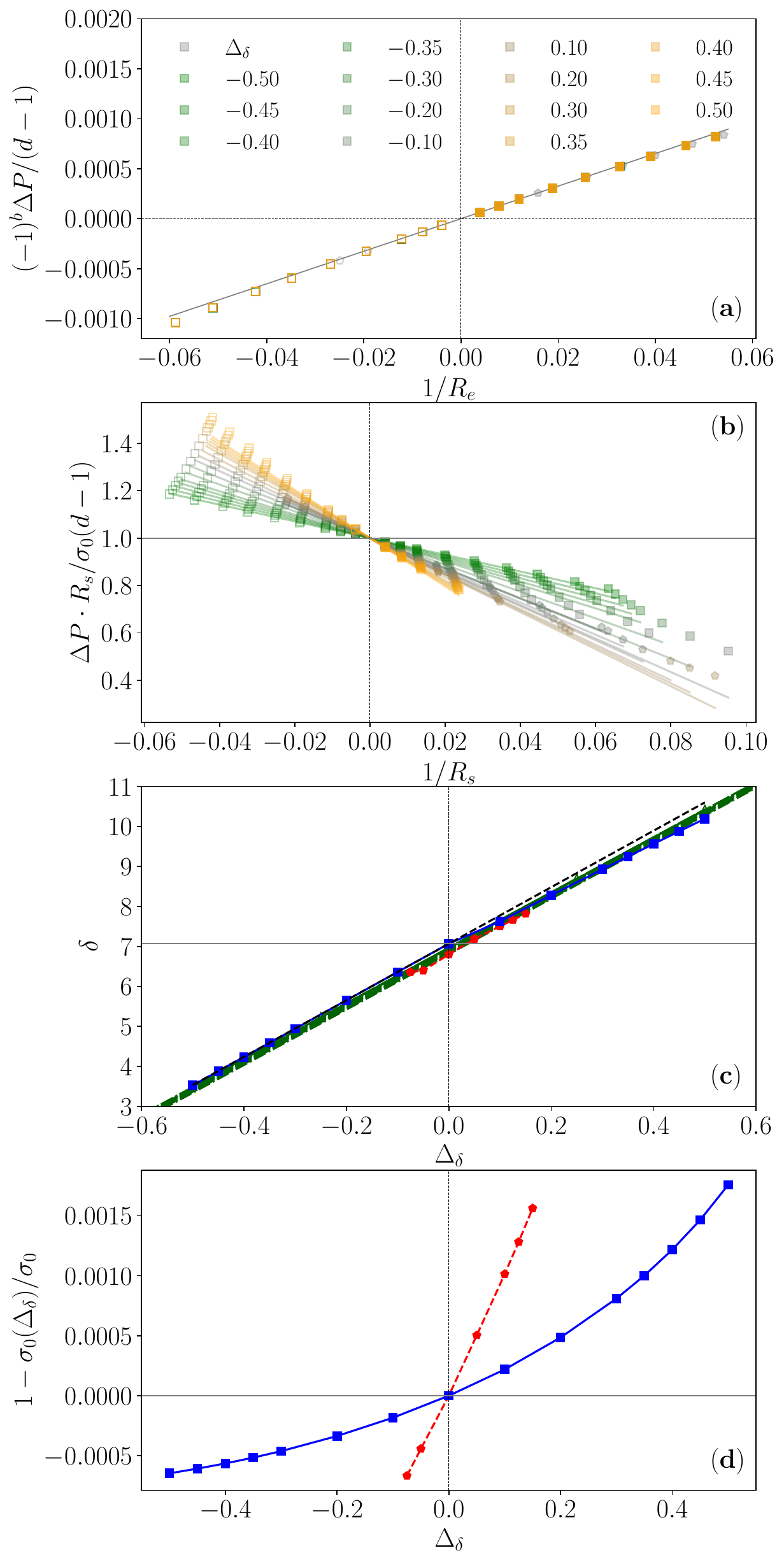}
\caption{Results for $\Delta\hat{T} = 0.1$ for curved interfaces in two and three dimensions. Negative curvatures, $R_e^{-1}<0$ or $R_s^{-1}<0$, are meant to indicate bubbles for which $b=1$. Panel $(a)$: Laplace law data for different values of the Tolman length tuning paramter $\Delta_\delta$. Panel $(b)$: corrections to the surface tension against $R_s^{-1}$ for different values of $\Delta_\delta$. Panel $(c)$: comparison of the estimation of the Tolman length from $d=1$ (empty triangle), $d=2$ (squares) and $d=3$ (pentagons) with expected tuning behavior in dashed black. Panel $(d)$: relative variation of the surface tension as estimated from the linear fit of the data in Panel $(b)$ around $R_s^{-1}\simeq0$.}\label{fig:closed0p1}
\end{figure}
\begin{figure}[!t]
\includegraphics[scale=0.46]{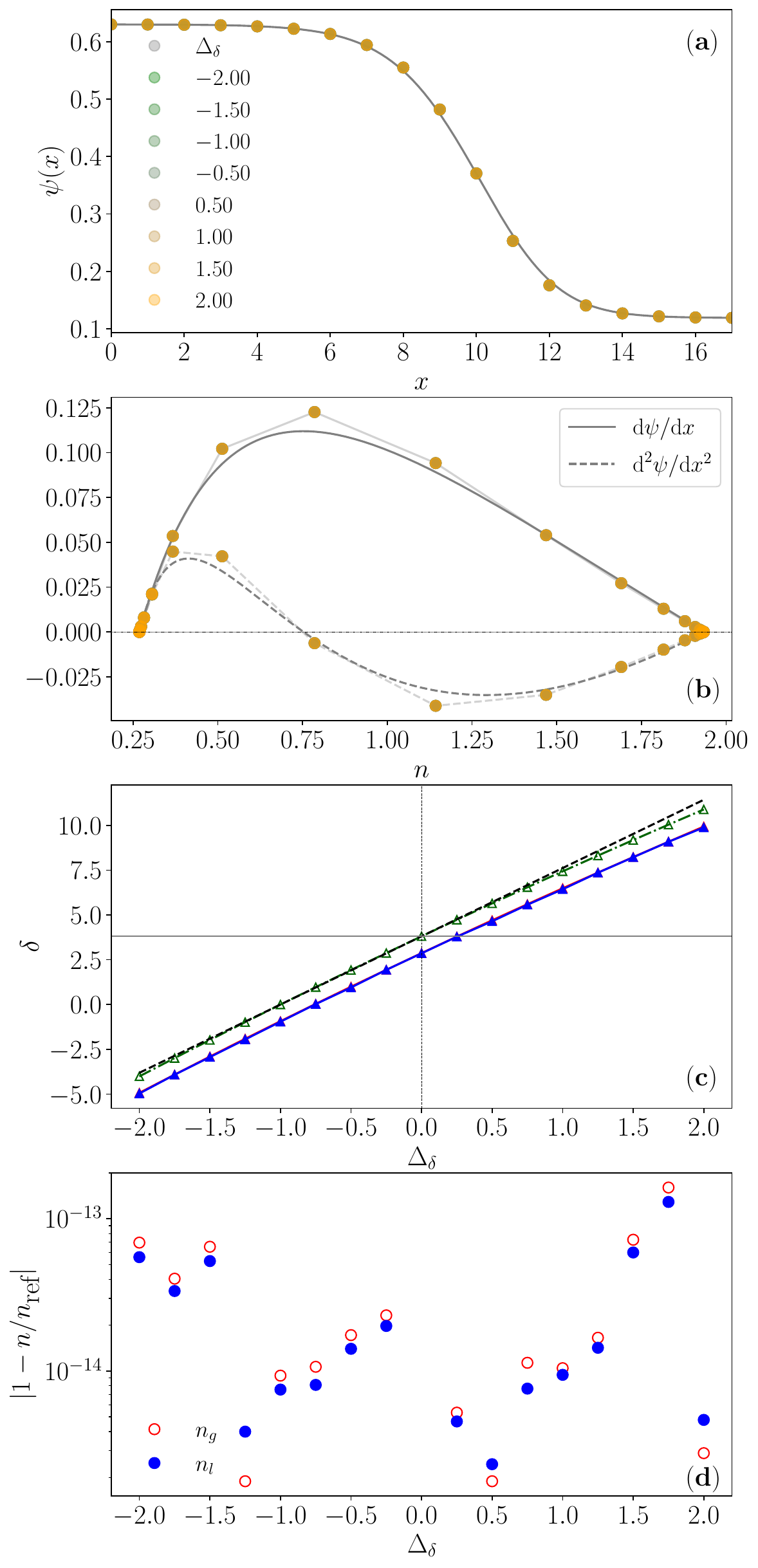}
\caption{Results for $\Delta\hat{T} = 0.2$ for flat interface simulations. Panel $(a)$: interface profile for several values of $\Delta_\delta$. Panel $(b)$: derivative profiles for the same values of $\Delta_\delta$ as in panel $(a)$. Panel $(c)$: different estimates of the Tolman length, according to Eq.~\eqref{eq:zemzs} and Eq.~\eqref{eq:deltaads} in red and blue triangles, and according to Eq.~\eqref{eq:SCTolmanFlat} in empty green triangles. The expected tuning behavior is reported in dashed black. Panel $(d)$: relative variation of the equilibrium bulk densities. Data in gray correspond to the reference value $\Delta_\delta=0$.}\label{fig:flat0p2}
\end{figure}
\begin{figure}[!t]
\includegraphics[scale=0.44]{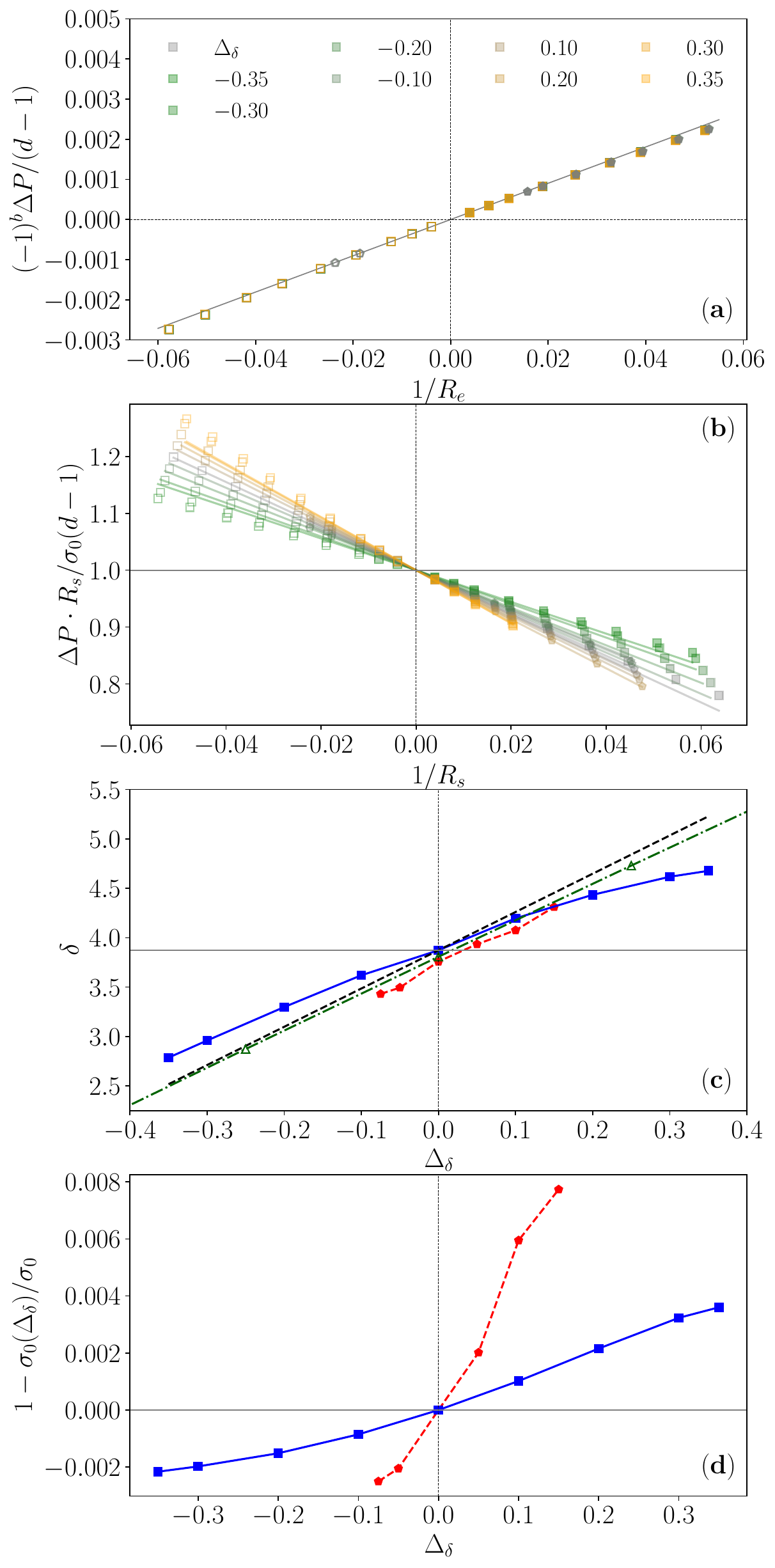}
\caption{Results for $\Delta\hat{T} = 0.2$ for curved interfaces in two and three dimensions. Negative curvatures, $R_e^{-1}<0$ or $R_s^{-1}<0$, are meant to indicate bubbles for which $b=1$. Panel $(a)$: Laplace law data for different values of the Tolman length tuning paramter $\Delta_\delta$. Panel $(b)$: corrections to the surface tension against $R_s^{-1}$ for different values of $\Delta_\delta$. Panel $(c)$: comparison of the estimation of the Tolman length from $d=1$ (empty triangle), $d=2$ (squares) and $d=3$ (pentagons) with expected tuning behavior in dashed black. Panel $(d)$: relative variation of the surface tension as estimated from the linear fit of the data in Panel $(b)$ around $R_s^{-1}\simeq0$.}\label{fig:closed0p2}
\end{figure}

In order to tune the Tolman length we need to introduce one more knob in the solution of the weights with respect to the previous subsection~\ref{ssec:s1}, namely $e_8$. This naturally comes to the cost of a lower degree of isotropy for the forcing, which would in turn yield stronger spurious currents. In order to remedy this issue we resort to the strategy described in the subsection~\ref{ssec:s0}, i.e., we choose a larger reference value for the fourth-order isotropy coefficient, i.e., $e_4=4/5$ and keep on using a dimensionful scaling for the higher-order ones, namely $e_6=e_4^{3/2}$ and $e_8=e_4^{2}$. Thermodynamic consistency is assured by means of the pseudo-potential defined in Eq.~\eqref{eq:psieps} for the corresponding value of $\varepsilon=14/29$ with $G_c=-2519424 \sqrt[7]{31888098}/12313081$. Within this setting the expression for the prefactor of the Tolman length can be written as
\begin{equation}
\hat{\delta}_{0}=\frac{1}{12}+\frac{95e_{4}}{432}+\frac{7e_{6}}{72}-\frac{5e_{8}}{144},
\label{eq:hatdelta0eI}
\end{equation}
which can be used, together with Eq.~\eqref{eq:hatsigma01eI}, to express the isotropy coefficients in terms of $\hat{\sigma}_0$, $\hat{\sigma}_1$ and $\hat{\delta}_0$. Given that the leading and the first order surface tension coefficients can be written as [see Eq.~\eqref{eq:hatsigma01eI}]
\begin{equation*}
\hat{\sigma}_{0}=-\frac{e_{4}}{2},\qquad\hat{\sigma}_{1}=\frac{e_{6}}{4},
\end{equation*}
 the tuning of the Tolman length can be achieved by keeping $\hat{\sigma}_0$ and $\hat{\sigma}_1$ fixed, and only changing $e_8$. Hence the expression of the Tolman length after the change $\hat{\delta}_0\to\hat{\delta}_0(1 + \Delta_\delta)$ simply reads
\begin{equation}
\delta(\Delta_\delta)\simeq\delta\cdot(1 + \Delta_\delta),
\end{equation}
with $\Delta_\delta = - 5\Delta e_8 / 144$, with $\Delta e_8$ the deviation from the reference value of the 8$^{th}$ order isotropy coefficient. All results in this subsection will be parametrized with respect to $\Delta_\delta$.


In order to test the robustness of the method we estimate the Tolman length in one dimension using Eq.~\eqref{eq:zemzs}, Eq.~\eqref{eq:SCTolmanFlat} and the following definition used in Tolman's work~\cite{Tolman1949}
\begin{equation}
\delta = \frac{\Gamma(z_s)}{n_l - n_g}.
\label{eq:deltaads}
\end{equation}
Results for the one-dimensional flat interface are reported in Fig.~\ref{fig:flat0p1} and Fig.~\ref{fig:flat0p2}: differently from what observed for the higher-order tuning of the surface tension in Section~\ref{ssec:s1}, the profiles shown in panels $(a)$ basically remain unchanged in a rather wide range of values of the tuning parameter $\Delta_\delta$, as well as the derivatives shown in panels $(b)$ displaying a good convergence for all the considered values of $\Delta_\delta$. As for the value of the Tolman length $\delta$ reported in panels $(c)$, the definitions from Eq.~\eqref{eq:zemzs} and Eq.~\eqref{eq:deltaads} yield identical results while the SC definition from Eq.~\eqref{eq:SCTolmanFlat} shows a roughly constant offset with respect to the latter. For both temperatures however, the relative change of $\delta$ correctly follows the expected $1 + \Delta_\delta$ linear behavior with excellent precision for $\Delta\hat{T}=0.1$ and with some slight deviation for $\Delta\hat{T}=0.2$. Finally, as reported in panels $(d)$, the relative change in the equilibrium bulk densities for the liquid $n_l$ and vapor $n_g$ is zero within machine precision, i.e., $O(10^{-13})$. In Fig.~\ref{fig:closed0p1} and Fig.~\ref{fig:closed0p2} we show the results for the case of bubbles and droplets in two and three dimensions for $\Delta\hat{T}=0.1$ and $\Delta\hat{T}=0.2$, respectively. The data reported in panels $(a)$ show that the Laplace law is correctly recovered at fixed slopes independently on the value of $\Delta_\delta$. In panels $(b)$ we report the Laplace law corrections as a function of $R_s^{-1}$, where the fanning of the data indicates the tuning of the Tolman length. In panels $(c)$ we report the estimated values of $\delta$ for all dimensions displaying an excellent matching with the analytical form of Eq.~\eqref{eq:SCTolmanFlat} for $\Delta\hat{T}=0.1$. At the lower temperature $\Delta\hat{T}=0.2$ the match is still very good. Finally, in panels $(d)$ we report the relative change of the surface tension, which is of order $10^{-3}$ for both temperatures. We notice that the values of $\delta$ collapse on the same curve independently on the dimension, indicating that the SC model yields a slightly different parametrization of the curvature corrections with respect to Eq.~\eqref{eq:SigmaCExpansion}.

In summary, the results for the tuning of the Tolman length are in excellent agreement with the theoretical estimates of (i) equilibrium bulk densities, (ii) first and second derivative of the interface profile, (iii) expected change in $\delta$ according to the SC definition of Eq.~\eqref{eq:SCTolmanFlat}. Indeed, the higher-order tuning strategy allows to fully control the surface tension, i.e., keeping it constant, while changing the value of the Tolman length in a rather wide range for $d=1$ and $d=2$, while the stable range is narrower for $d=3$. We remark that, in the present setting, it is only possible to \emph{change the sign} of the Tolman length in the one-dimensional case, while simulations become unstable for very large values of $\Delta_\delta$ in two and three dimensions. However, the inclusion of $F^2$ terms in the forcing, along the direction of~\cite{Li_2013,khajepor2016multipseudopotential}, should allow for more stable simulations, possibly allowing for a sign change for closed interfaces.

\begin{figure*}[!t]
\includegraphics[scale=0.425]{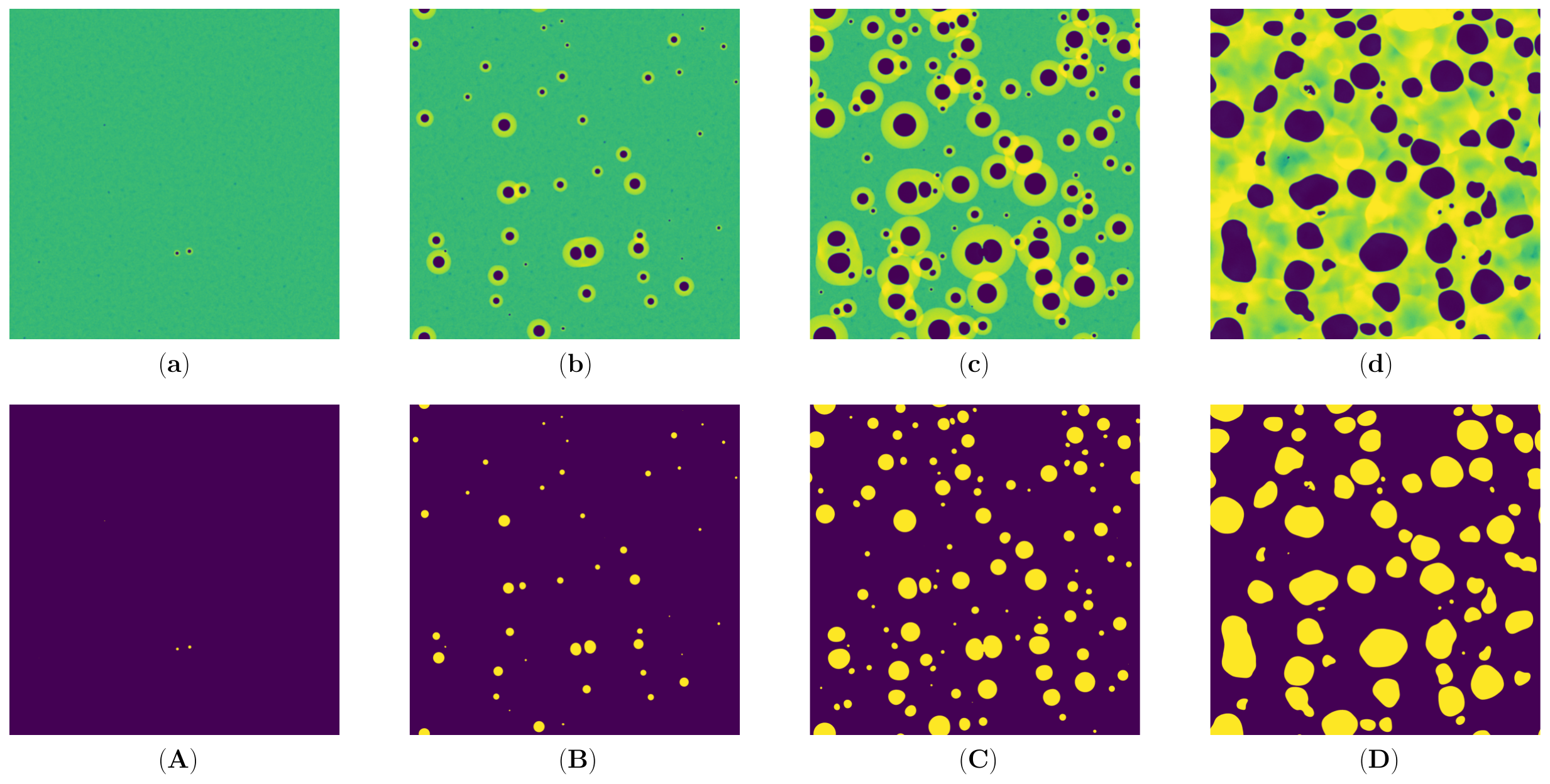}
\caption{Panels displaying the different stages of homogeneous nucleation dynamics for $\Delta_\delta = 0$ using the density field (upper row) and a binary-mask field (lower row) used for counting the number of bubbles by means of clustering algorithms. From left to right the stages are $(a)$ nucleation inception $(t=2560)$, $(b)$ steady-state nucleation $(t=3840)$, $(c)$ onset of the interactions among independent nucleation events ($t=5120$) and $(d)$ coalescence of the largest bubbles $(t=9920)$. The time stamps in the video included in the Supplemental Material are $t=3.6$s, $t=5.2$s, $t=6.8$s and $t=12.8$s, respectively. Simulations are run for a square system of size $L=4096$ with an initial homogeneous liquid density $n_l=1.031$, reduced temperature $\Delta \hat{T}=0.056$ and hydrodynamic fluctuation temperature $k_BT=5\times 10^{-5}$.}\label{fig:nuc_rate_panel}
\end{figure*}
\subsection{Nucleation Rates}
Finally, we report some fluctuating hydrodynamic results demonstrating the dependence of the homogeneous nucleation rate $J$ on the Tolman length at a fixed value of the surface tension. The forcing stencil is the same as in the previous subsection with $\Delta \hat{T}=0.056$, i.e., closer to the critical point. We use the implementation for fluctuating hydrodynamics in the SC multiphase model described in~\cite{lulli2024metastable}. We set the hydrodynamic temperature to $k_BT=5\times 10^{-5}$ and initialize the system in a metastable state: the homogeneous initial density value is set according to $$n_{\text{init}}(a)=n_{s,l} (1 - a) + an_l$$ with $a\simeq0.02$, where $n_l\simeq 1.264$ is the equilibrium liquid density and $n_{s,l}\simeq1.026$ is the density at the spinodal point of the liquid branch of the equation of state. Such values are chosen for the nucleation rate to be large enough and allow for relatively short simulations. In order to measure the steady-state nucleation rate we use a square system of linear size $L=4096$ lattice sites, such that a statistically significant number of independent nucleation events can occur. Given that our setup exactly conserves mass, the nucleation dynamics will evolve in four different phases depicted in Fig.~\ref{fig:nuc_rate_panel}: (i) initial stages of nucleation with only a few nucleating embryos as in Fig.~\ref{fig:nuc_rate_panel}$(a)$, (ii) \emph{steady-state} regime where independent nucleation events occur in the system as in Fig.~\ref{fig:nuc_rate_panel}$(b)$, (iii) the number of nucleating bubbles reaches a maximum as different nucleation events interact with each other as in Fig.~\ref{fig:nuc_rate_panel}$(c)$ and (iv) the number of bubbles decreases with the largest bubbles coalescing as in Fig.~\ref{fig:nuc_rate_panel}$(d)$. Importantly, in Fig.~\ref{fig:nuc_rate_panel}$(b)$ one can see that in the steady-state regime nucleation events are independent because around each nucleating bubble there is a higher density halo appearing in lighter color, which can be interpreted as a sound wave, and most nucleation events have non-overlapping halos, differently from Fig.~\ref{fig:nuc_rate_panel}$(c)$. In the lower panels in Fig.~\ref{fig:nuc_rate_panel} we also report the ``binary mask'' field used for counting the number of bubbles via standard clustering algorithms available in the package scipy~\cite{scipy}. The binary field is obtained by setting to ``1'' the sites such that the difference between the equilibrium liquid density and the local density is larger than $3/4$ the difference between liquid and gas densities at equilibrium, while the other sites are set to ``0''.
\begin{figure}[!t]
\includegraphics[scale=0.425]{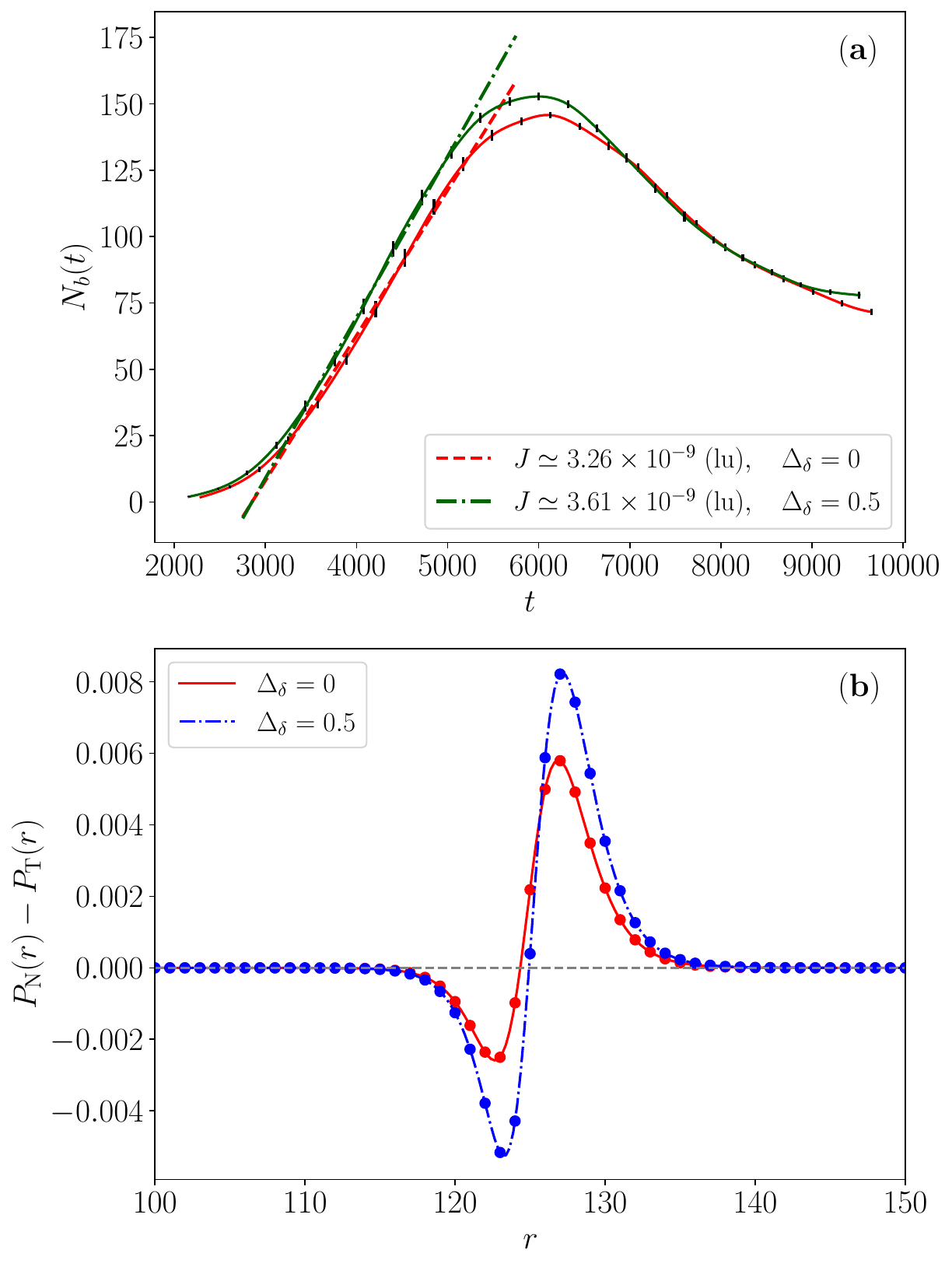}
\caption{Panel $(a)$: number of nucleating bubbles as a function of time $N_b(t)$ in a fully periodic system with $L=4096$. Each curve represents the average over $N_{\text{rep}} = 32$ independent simulations. The steady-state nucleation rates are estimated from the linear fits indicated by the dashed lines. Error bars are reported every ten data points. Panel $(b)$: Radial profiles for $P_{\text{N}} - P_{\text{T}}$ for the two different values of $\Delta_\delta$ for stationary bubbles of size $R\simeq128$(lu).}\label{fig:nuc_rate_Nb}
\end{figure}

The four stages described above can be recognized as well in Fig.~\ref{fig:nuc_rate_Nb}$(a)$ displaying the number of bubbles as a function of time $N_b(t)$, starting from zero, then growing until a maximum is reached, followed by a monotonic decrease. We study the evolution of this quantity for two different tuning values of the Tolman length $\Delta_\delta = 0$ and $\Delta_\delta=0.5$, with the reference value $\hat{\delta}_0=\frac{7 \sqrt{5}}{225} + \frac{32}{135}\simeq0.307$ and $\delta(\Delta_\delta) \simeq 10.97(1 + \Delta_\delta)$ (lu). Given the curvature dependence of the surface tension and that of the free-energy barrier, i.e., $$\sigma\left(R_{s}\right)\simeq\sigma_{0}\left(1-\frac{\delta}{R_{s}}\right),\quad \Delta F\approx-4\pi R_{c}^{2}\sigma(R_s),$$ we expect a decrease of $\Delta F$ as $\delta$ increases, i.e., larger nucleation rates for $\Delta_\delta > 0$ since $$J \propto \exp\left[-\frac{\Delta F}{k_B T}\right].$$ This can be observed in Fig.~\ref{fig:nuc_rate_Nb}$(a)$ displaying a larger slope for the case $\Delta_\delta=0.5$. After normalizing the slopes by the system area one obtains the nucleation rates $J$ which in this case vary by roughly $10\%$ as a response to changing the Tolman length $\delta$ while keeping the flat surface tension $\sigma_0$ unchanged, as it can be seen from the radial profiles of $P_{\text{N}} - P_{\text{T}}$ in Fig.~\ref{fig:nuc_rate_Nb}$(b)$ for stationary bubbles of size $R\simeq 128$(lu), for both values of $\Delta_\delta$. These results demonstrate the applicability of the present work to more complex hydrodynamic regimes as well as the possibility for the SC multiphase LBM to model nucleation physics beyond the CNT framework. In the supplemental material we provide a video displaying more clearly all the features of the nucleation dynamics in the different stages.

\section{Discussion \& Conclusions}\label{sec:dandc}
When using the SC model for multiphase flows, several strategies have been devised to independently tune different physical properties of a multiphase mixture. In the seminal work~\cite{Sbragaglia_2007} an extension of the SC model was proposed that was aiming at tuning separately the density ratio and the surface tension: this was achieved by using two copies of the interaction potential, one for short- and one for mid-range interactions, i.e., a \emph{multirange} approach, which were connected to two different coupling constants. By means of two independent linear combinations it was possible to independently tune the effective temperature and the prefactor to the integral appearing in the expression of the surface tension. In the works~\cite{Falcucci_2007,Chibbaro_2008,Falcucci_2007} it was proposed to use two different interactions potentials, i.e., a \emph{multibelt} approach, that would be again associated with two different coupling constants allowing for the separate tuning of the effective temperature and surface tension prefactor. Both the multibelt and multirange approaches have led to important advances in the field. Most notably, the multibelt approach was used in~\cite{Chibbaro_2008}, in conjunction with the SC implementation of the forcing~\cite{Shan_1993}, in order to obtain a negative surface tension at large scales, inducing an emulsion-like behaviour in a multiphase system, while in~\cite{Benzi_2009} the multirange approach was used in a multicomponent framework in order to model emulsions with a small but positive surface tension allowing for a direct comparison with experiments~\cite{Derzsi_2017,Derzsi_2018,Pelusi_2022,pelusi2024emulsions}. Moreover, the multibelt approach has been successfully used to model cavitation inception in realistic geometries and loading schedules~\cite{Falcucci_2013,Falcucci_2013_1}. However, in~\cite{Li_2013} it was noticed that neither the multirange nor the multibelt approach was able to completely decouple the density ratio from the surface tension: a new LBM implementation was devised, leveraging the lattice pressure tensor~\cite{Shan_2008} and the multiple-relaxation-time (MRT) collision operator~\cite{Kruger_2017,succi2018lattice}, in order to tune independently the flat-interface profile and the surface tension. Starting from these results, yet another approach, labelled as \emph{multipseudopotential}~\cite{Khajepor_2015,khajepor2016multipseudopotential}, was proposed in order to tune separately various interface properties, such as density ratio, interface width and surface tension. This result was achieved by considering a body force obtained by the combination of forces computed with different pseudopotential functions $\{\psi_i\}$ as well as different coupling constants and a specific modification of Guo's forcing scheme~\cite{Guo_2002}. It is worthwhile noticing that the full decoupling of flat-interface profile and surface tension has only been obtained in~\cite{Li_2013,Khajepor_2015,khajepor2016multipseudopotential} by means of rather extensive modifications of the original model, either by resorting to the MRT approach or by introducing extra forcing terms along with multiple body-force components. 

Our work presents a different approach allowing for the tuning of the interface physics, both at leading and higher-order, for fixed bulk properties. In particular, we demonstrated for the first time the ability to tune the value of the Tolman length while keeping fixed surface tension, bulk densities and flat interface profiles. This is achieved by extending the usual equilibrium analysis to include second-order derivatives of the density profile and by leveraging \emph{all} the degrees of freedom hidden in the SC forcing stencil weights~\cite{Shan_1994}. The latter are unlocked by requiring a non-maximal degree of isotropy for any given stencil. In turn, this enables us to write the weights in terms of physical knobs thus providing a direct interpretation of their values while always assuring a constant and highest degree of isotropy. In doing so we provide (i) a set of weights reducing the spurious currents while keeping a 6$^{th}$ order isotropy (see Table~\ref{tab:e4_tuning_6}), (ii) a set of weights for tuning the surface tension by means of higher-order effects while keeping the bulk densities constant and (iii) a set of weights for tuning the Tolman length (see Tables~\ref{tab:e4e6e8_tuning_6_2d} and~\ref{tab:e4e6e8_tuning_6_3d}). These results rely on the knowledge of the SC lattice pressure tensor~\cite{Shan_2008,Lulli_2021} which we have used to derive the first analytical expression for the Tolman length in the SC multiphase model [see Eq.~\eqref{eq:SCTolmanFlat}], and our analytical result matches the numerical results in two and three dimensions with great accuracy [see Fig.~\ref{fig:closed0p1}$(c)$ and Fig.~\ref{fig:closed0p2}$(c)$]. Given the relevance of higher-order interface physics in nucleation/cavitation processes, we use the present approach in conjunction with a recent implementation of fluctuating hydrodynamics for the SC multiphase model~\cite{lulli2024metastable}, to demonstrate that homogeneous nucleation rates are sensitive to the value of the Tolman length for a constant value of the surface tension. To the best of our knowledge, this is the first instance that such a control is realized in the context of multiphase LBM, demonstrating the ability of the current approach to model nucleation processes beyond the usual CNT approximation. We remark that all our results have been obtained in one of the simplest and most straightforward LBM framework, i.e., the single-relaxation-time BGK with Guo forcing~\cite{Guo_2002} implementation. 

From a more general perspective one can aim at leveraging the strategy proposed in this work for the tuning of properties other than the Tolman length or the surface tension. As shown in~\cite{lulli2024metastable} it is possible to express some of the coefficients in the Fourier expansion of the structure factor $S(k)$ as a linear combination of the stencil moments; at the same time, $S(k)$ can also be expressed in terms of the radial distribution function $g(r)$, which is known to be connected to the microscopic atomic/molecular interaction details through the \emph{exponential of mean force} $w^{(2)}(r)$ defined as the marginalization of the interaction potential $V=V(r_1,r_2,\ldots,r_n)$ over the coordinates of all particles but a pair. Hence, provided enough degrees of freedom are available in the stencil, the present approach could allow for a mesoscopic modelling of the interface properties that are also compatible with the microscopic potential interactions of real substances.

We briefly comment on the numerical cost associated with the method proposed in this work: the only overhead with respect to the vanilla LBM implementation with Guo forcing scheme~\cite{Guo_2002} lies in the use of a wider interaction stencil, implying a larger number of memory reads. This has been dealt with effectively by resorting to automatic loop unrolling in the metaprogramming approach used to perform the simulations of the present work using the idea.deploy framework. In other words, all the heavy lifting is made by the appropriate choice of the weights that are provided in the Tables~\ref{tab:e4_tuning_6},~\ref{tab:e4e6_tuning_8},~\ref{tab:e4e6e8_tuning_6_2d} and~\ref{tab:e4e6e8_tuning_6_3d} explicitly in terms of the physical knobs of interest. A different strategy has been used in~\cite{Khajepor_2015,khajepor2016multipseudopotential} and~\cite{Li_2013} where multiple instances of different pseudo-potential function $\psi_i$ have been used as well as the MRT approach; while these changes might not immediately impact memory bandwidth, they are more computationally intensive.

The source code for the simulations can be found on the github repository \href{https://github.com/lullimat/idea.deploy}{https://github.com/lullimat/idea.deploy}~\cite{sympy, scipy, numpy0, numpy1, scikit-learn, matplotlib, ipython, pycuda_opencl}, where a Jupyter notebook~\cite{ipython} is available to reproduce all the results reported in this paper.

In summary, we demonstrate the efficacy of the present method for using the SC model in the study of out-of-equilibrium phase-changing phenomena capturing the role of the microscopic interface physics. A combination of these techniques, possibly in conjunction with suitable modification of the unstable part of the equation of state~\cite{colosqui2012mesoscopic,czelusniak2022shaping}, and the introduction of $F^2$ terms and the multi-pseudopotential approach~\cite{Li_2013,Khajepor_2015,khajepor2016multipseudopotential} would pave the way for a high-fidelity modelling of complex non-equilibrium hydrodynamics of real substances.

\begin{acknowledgments}
The authors wish to thank Xiaowen Shan, Mauro Sbragaglia, Luca Biferale, Giacomo Falcucci and Dong Yang for useful and fruitful discussions.
\end{acknowledgments}

\appendix
\section{Stencil Isotropy Example: $E^{(4)}$}\label{app:e4ex}
The simplest example of this is provided by the two-dimensional forcing stencil with vectors chosen as those in Fig.~\ref{fig:stencil}$(a)$, i.e., two symmetry groups of squared lengths $\ell^2=|\mathbf{e}_a|^2=1,2$ for which one can consider at most a fourth order isotropy of the moments
\begin{equation}
\begin{split}\langle e^{\mu}e^{\nu}\rangle_{W} & =e_{2}\delta^{\mu\nu}\\
\langle e^{\mu}e^{\nu}e^{\rho}e^{\sigma}\rangle_{W} & =e_{4}\Delta^{\mu\nu\rho\sigma}+I_{4,0}\delta^{\mu\nu\rho\sigma}.
\end{split}
\end{equation}
The above relations can be used to determine $e_2, e_4$ and $I_{4,0}$ as a function of the weights $W(1)$ and $W(2)$
\begin{equation}
\begin{split}e_{2} & =2W\left(1\right)+4W\left(2\right),\qquad e_{4}=4W\left(2\right),\\
I_{4,0} & =2W\left(1\right)-8W\left(2\right).
\end{split}
\end{equation}
It is then possible to express the weights using only two of the above: given the importance of the forcing isotropy we select $e_2$ and $I_{4,0}$, yielding
\begin{equation}
\begin{split}W\left(1\right) & =\frac{1}{6}\left(2e_{2}+I_{4,0}\right),\\
W\left(2\right) & =\frac{1}{12}\left(e_{2}-I_{4,0}\right).
\end{split}
\end{equation}
Selecting any non-vanishing value for the anisotropy coefficient $I_{4,0}\neq 0$ would yield to severe artifacts such as square shaped droplets and large spurious currents. Hence, when selecting $I_{4,0}=0$ and a unit variance $e_2=1$ one recovers the usual values of the weights, namely $W(1)=1/3$ and $W(2)=1/12$~\cite{Shan_2008}.

\section{Recovering the Multirange \& Multibelt Approaches}\label{app:MBMR}
At this point it is interesting to draw an explicit connection between the present approach, the multirange~\cite{Sbragaglia_2007,Benzi_2009} and the multibelt works~\cite{Falcucci_2007,Chibbaro_2008}. According to the multirange paradigm a generalized Shan-Chen force $F^\mu_\text{MR}$ is defined as
\begin{equation}
F_{\text{MR}}^{\mu}\left(\mathbf{x}\right)=-c_{s}^{2}\psi\left(\mathbf{x}\right)\langle\left[G_{1}\psi\left(\mathbf{x}+\mathbf{e}\right)+G_{2}\psi\left(\mathbf{x}+2\mathbf{e}\right)\right]e^{\mu}\rangle_{W}
\end{equation}
where the underlying stencil is chosen to be the fourth-order isotropic one. It is possible to split the expression above in two contributions belonging to two different stencils
\begin{equation}
\begin{split}
&F_{\text{MR}}^{\mu}\left(\mathbf{x}\right)=-G_{1}c_{s}^{2}\psi\left(\mathbf{x}\right)\\&\qquad\times\left(\langle\psi\left(\mathbf{x}+\mathbf{e}\right)e^{\mu}\rangle_{W}+g_{\text{21}}\langle\psi\left(\mathbf{x}+\bar{\mathbf{e}}\right)\bar{e}^{\mu}\rangle_{\bar{W}}\right)
\end{split}
\end{equation}
where $g_{21}=G_2 / G_1$, the new vectors are $|\bar{\mathbf{e}}_a|^2\in\{4,8\}$ and the new weights can be expressed in terms of the original ones as $\bar{W}(4) = W(1) / 2 = 1/6$ and $\bar{W}(8) = W(4) / 2 = 1/24$. The new stencil is still fourth-order isotropic (see general expression for $I_{4,0}$ in~\cite{Lulli_2021}) since
\begin{equation}
\begin{split} & \bar{I}_{4,0}=32\bar{W}\left(4\right)-128\bar{W}\left(8\right)\\
 & \qquad=8\left[2W\left(1\right)-8W\left(2\right)\right]=2I_{4,0}=0,
\end{split}
\end{equation}
and the first two isotropy coefficients read
\begin{equation}
\begin{split}
&\bar{e}_{2}=8\bar{W}\left(4\right)+16\bar{W}\left(8\right)=2,\\&\bar{e}_{4}=64\bar{W}\left(8\right)=\frac{8}{3}.
\end{split}
\end{equation}
In the present framework this is equivalent to defining a stencil with $N_\text{W}=4$ degrees of freedom according to the conditions
\begin{equation}
\begin{split}
&\hat{e}_{2}=e_{2}+g_{21}\bar{e}_{2}=1+2g_{21},\\&\hat{e}_{4}=e_{4}+g_{21}\bar{e}_{4}=\frac{1}{3}\left(1+8g_{12}\right),\\&\hat{e}_{6}=e_{6}+g_{21}\bar{e}_{6},\qquad\hat{I}_{4,0}=0.
\end{split}
\end{equation}
Indeed, the first two equalities above, match the definitions given in Eq.(16) of~\cite{Sbragaglia_2007}, e.g. $A_1 = G_1 \hat{e}_2$ and $A_2 c_s^2 = G_1 \hat{e}_4$, representing the two main physical \emph{knobs} tuning the interaction strength and surface tension respectively. The equivalent stencil reads
\begin{equation}
\begin{split}
&\hat{W}\left(1\right)=\frac{1}{3},\quad\hat{W}\left(2\right)=\frac{1}{12},\\&\hat{W}\left(4\right)=\frac{g_{21}}{6},\quad\hat{W}\left(8\right)=\frac{g_{21}}{24},
\end{split}
\end{equation}
so that the Shan-Chen force can be written in the usual way
\begin{equation}
F^{\mu}\left(\mathbf{x}\right)=-Gc_{s}^{2}\psi\left(\mathbf{x}\right)\langle\psi\left(\mathbf{x}+\mathbf{e}\right)e^{\mu}\rangle_{\hat{W}}.
\end{equation}
As expected the overall stencil is just the juxtaposition of the previous two. From the present perspective, the multirange scheme has been devised as to tune independently $e_2$ and $e_4$ while keeping fixed the isotropy condition $I_{4,0}$. However, we are using a total of $N_\text{W}=4$ degrees of freedom rather than three, since the scheme is implicitly tuning also $e_6$. Indeed, given that $e_6$ is not associated with any physical observable in the approximation used in~\cite{Sbragaglia_2007} it follows that a possible improvement over the scheme would be to trade the tuning of $e_6$ for the sixth-order isotropy condition of the forcing. This can be achieved by requiring the weights to satisfy the system of equations
\begin{equation}
\begin{split}
&\hat{e}_{2}=e_{2}+g_{21}\bar{e}_{2}=1+2g_{21},\\&\hat{e}_{4}=e_{4}+g_{21}\bar{e}_{4}=\frac{1}{3}\left(1+8g_{12}\right),\\&\hat{I}_{4,0}=0,\qquad\hat{I}_{6,0}=0.
\end{split}
\end{equation}
which in turn yields the set of weights satisfying the fourth- and sixth-order isotropy condition for all values of $g_{21}$
\begin{equation}
\begin{split}
&\hat{W}\left(1\right)=\frac{14}{45}-\frac{32g_{12}}{45},\quad\hat{W}\left(2\right)=\frac{16g_{12}}{45}+\frac{17}{180},\\&\hat{W}\left(4\right)=\frac{19g_{12}}{90}+\frac{1}{720},\quad\hat{W}\left(8\right)=\frac{7g_{12}}{360}-\frac{1}{1440}.
\end{split}
\end{equation}

Next, let us consider the multibelt approach described in~\cite{Falcucci_2007,Chibbaro_2008}: this case is more straightforward as one starts by defining a generalized Shan-Chen forcing directly defined as the sum of two contributions coming from two different stencils
\begin{equation}
\begin{split}
&F_{\text{MB}}^{\mu}\left(\mathbf{x}\right)=-c_{s}^{2}\psi\left(\mathbf{x}\right)\\&\qquad\times\left(G_{1}\langle\psi\left(\mathbf{x}+\mathbf{e}\right)e^{\mu}\rangle_{W}+G_{2}\langle\psi\left(\mathbf{x}+\bar{\mathbf{e}}\right)\bar{e}^{\mu}\rangle_{\bar{W}}\right),
\end{split}
\end{equation}
where the first term corresponds once more to the fourth-order isotropic stencil while the second term corresponds to the eighth-order isotropic stencil with weights $\bar{W}(1)=4/21$, $\bar{W}(2)=4/45$, $\bar{W}(4)=1/60$, $\bar{W}(5)=2/315$ and $\bar{W}(8)=1/5040$. Similarly to the multirange case, it is possible to define a unique stencil respecting all the relevant properties by solving the system of equations
\begin{equation}
\begin{split}
&\hat{e}_{2}=e_{2}+g_{21}\bar{e}_{2}=1+g_{21},\\&\hat{e}_{4}=e_{4}+g_{21}\bar{e}_{4}=\frac{1}{3}\left(1+\frac{12}{7}g_{21}\right),\\&\hat{e}_{6}=e_{6}+g_{21}\bar{e}_{6},\\&\hat{e}_{8}=e_{8}+g_{21}\bar{e}_{8},\quad\hat{I}_{4,0}=0.
\end{split}
\end{equation}
The first two lines correspond to the coefficients reported in Eq. (14) and (15) of~\cite{Chibbaro_2008}. The resulting set of weights reads
\begin{equation}
\begin{split}
&\hat{W}\left(1\right)=\frac{1}{3}+\frac{4g_{12}}{21},\quad\hat{W}\left(2\right)=\frac{1}{12}+\frac{4g_{12}}{45},\\&\hat{W}\left(4\right)=\frac{g_{12}}{60},\quad\hat{W}\left(5\right)=\frac{2g_{12}}{315},\\&\hat{W}\left(8\right)=\frac{g_{12}}{5040},
\end{split}
\end{equation}
where we can clearly recognize again the linear combination of the two stencils as expected. However, we are using $N_\text{W}=5$ degrees of freedom for tuning only two physical quantities, namely $\hat{e}_2$ and $\hat{e}_4$, while there is no physical observable associated with $\hat{e}_6$ and $\hat{e}_8$. Hence, one could replace those conditions and $\hat{I}_{4,0}$ with the fourth-order isotropy conditions for the lattice pressure tensor~\cite{Lulli_2021} $\chi_I$ and $\Lambda_I$ and with the sixth-order isotropy condition for the forcing $I_{6,0}$ yielding the weights
\begin{equation}
\begin{split}
&\hat{W}\left(1\right)=\frac{4g_{12}}{15}+\frac{121}{360},\quad\hat{W}\left(2\right)=\frac{4g_{12}}{105}+\frac{7}{90},\\&\hat{W}\left(4\right)=-\frac{g_{12}}{420}-\frac{7}{1440},\quad\hat{W}\left(5\right)=\frac{2g_{12}}{105}+\frac{1}{240},\\&\hat{W}\left(8\right)=-\frac{g_{12}}{336}-\frac{1}{576},
\end{split}
\end{equation}
which satisfies the conditions $\chi_I = \Lambda_I = I_{6,0} = 0$ for all values of $g_{21}$. In summary, for both the multirange~\cite{Sbragaglia_2007,Benzi_2009} and the multibelt~\cite{Falcucci_2007,Chibbaro_2008} approaches, we managed to (i) consistently define a system of equations involving the stencil moments and derive the solution and to (ii) improve both forcing schemes with the introduction of higher-order isotropy conditions after having recognized that not all the available degrees of freedom were used. This is the main advantage of the present perspective: to provide a clear way to identify the available degrees of freedom of the forcing stencil and determine the weights directly from imposing conditions on the stencil moments, i.e., isotropy and anisotropy coefficients.

\renewcommand{\arraystretch}{2.3}
\begin{table*}[t!]
  \begin{ruledtabular}
  \centering
  \begin{tabular}{p{7mm}p{180mm}}
  \multicolumn{2}{c}{Two-dimensional isotropy coefficients} \\
  \hline
  $\displaystyle e_2$ & 
  $\displaystyle 2W(1)+4W(2)+8W(4)+20W(5)+16W(8)+18W(9)+32W(16)+36W(18)$\\
  
  $\displaystyle e_4$ & 
  $\displaystyle 4W(2)+32W(5)+64W(8)+324W(18)$ \\

  $\displaystyle e_6$ & 
  $\displaystyle \frac{4}{3}W(2)+\frac{80}{3}W(5)+\frac{256}{3}W(8)+972W(18)$ \\

  $\displaystyle e_8$ & 
  $\displaystyle \frac{4}{9}W(2)+\frac{128}{9}W(5)+\frac{1024}{9}W(8)+2916W(18)$ \\

  \hline
  \multicolumn{2}{c}{Two-dimensional anisotropy coefficients} \\
  \hline
  $\displaystyle I_{4,0}$ & $\displaystyle 2W(1)-8W(2)+32W(4)-28W(5)-128W(8)+162W(9)-648W(18)$\\
  $\displaystyle I_{6,0}$ & $\displaystyle 2W(1)-16W(2)+128W(4)-140W(5)-1024W(8)+1458W(9)-11664W(18)$\\
  $\displaystyle I_{8,0}$ & $\displaystyle -\frac{8}{3}W(2)+\frac{176}{3}W(5)-\frac{2048}{3}W(8)-17496W(18)$\\
  $\displaystyle I_{8,1}$ & $\displaystyle 2W(1)-157464W(18)-24W(2)+512W(4)-876W(5)-6144W(8)+13122W(9)$\\

  \hline
  \multicolumn{2}{c}{Three-dimensional anisotropy coefficients} \\
  \hline
  $\displaystyle I_{4,0}$ & $\displaystyle 2W(1)-4W(2)-16W(3)+32W(4)+40W(5)-72W(6)-64W(8)-312W_{(2,2,1)}(9)+162W_{(3,0,0)}(9)$\newline$+512W(16)-324W(18)$\\
  $\displaystyle I_{6,0}$ & $\displaystyle 2W(1)-52W(2)+128W(3)+128W(4)-680W(5)+888W(6)-3328W(8)+2472W_{(2,2,1)}(9)+1458W_{(3,0,0)}(9)$\newline$+8192W(16)-37908W(18)$\\
  $\displaystyle I_{6,1}$ & $\displaystyle 4W(2)-16W(3)+80W(5)-120W(6)+256W(8)-480W_{(2,2,1)}(9)+2916W_{(3,3,0)}(9)$\\
  $\displaystyle I_{8,0}$ & $\displaystyle 2W(1)-20W(2)-48W(3)+512W(4)+152W(5)-1800W(6)-5120W(8)-14040W_{(2,2,1)}(9)+13122W_{(3,0,0)}(9)$\newline$+131072W(16)-131220W(18)$\\
  $\displaystyle I_{8,1}$ & $\displaystyle -\frac{8}{3}W(2)-\frac{16}{3}W(3)+\frac{176}{3}W(5)+112W(6)-\frac{2048}{3}W(8)-1248W_{(2,2,1)}(9)-17496W(18)$\\
  $\displaystyle I_{8,2}$ & $\displaystyle 4W(2)-16W(3)+128W(5)-312W(6)+1024W(8)-1152W_{(2,2,1)}(9)+26244W(18)$\\
  $\displaystyle I_{10,0}$ & $\displaystyle 2W(1)-28W(2)-64W(3)+2048W(4)-1160W(5)-10584W(6)-28672W(8)-76056W_{(2,2,1)}(9)+118098W_{(3,0,0)}(9)$\newline$+2097152W(16)-1653372W(18)$\\
  $\displaystyle I_{10,1}$ & $\displaystyle -\frac{16}{3}W(2)-\frac{32}{3}W(3)+\frac{880}{3}W(5)+576W(6)-\frac{16384}{3}W(8)-10336W_{(2,2,1)}(9)-314928W(18)$\\
  $\displaystyle I_{10,2}$ & $\displaystyle 4W(2)-32W(3)+320W(5)-792W(6)+4096W(8)-6528W_{(2,2,1)}(9)+236196W(18)$\\
  \end{tabular}
  \end{ruledtabular}\caption{}\label{tab:defe2nI2n}
\end{table*}
\section{On the role of $\varepsilon$ and $e_4$}\label{app:epse4}
We now express $\varepsilon$ in terms of the isotropy coefficients~\cite{Shan_2008,Lulli_2021} which results in a function of $e_4$ only (since we always set $e_2=1$)
\begin{equation}
\varepsilon=\frac{6e_{4}-2e_{2}}{6e_{4}+e_{2}},
\label{eq:eps_moments}
\end{equation}
which converges to unity for very large $e_4$ and is equal to zero for the case with $e_4=1/3$, i.e., the fourth-order isotropic forcing scheme derived above. Indeed, $\varepsilon(e_4)$ is an invertible function in this domain so that $\varepsilon$ and $e_4$ can be understood as representing the same degree of freedom. As it can be seen from Eq.~\eqref{eq:SCMaxwell} and Eq.~\eqref{eq:PSIProfile} $\varepsilon$ plays a role both in the determination of the equilibrium densities and pressure, as well as in deciding the slope of the profile. In the last case the coefficient $(1-\varepsilon)$ makes it so that a larger value of $\varepsilon$, i.e., a larger value of $e_4$, would have a smaller slope yielding a smoother profile. On the other hand, one can also express the combination of coefficients appearing in the expression for the flat-interface surface tension as follows
\begin{equation}
c_{11}-c_{02}=-\frac{I_{4,0}}{4}-\frac{e_{4}}{2}=-\frac{e_{4}}{2},
\label{eq:hat_sigma_0_moments}
\end{equation}
where in the last equality we have imposed the fourth-order isotropy by imposing $I_{4,0}=0$. This is a well-known result~\cite{Shan_2008,Lulli_2021}. However, it has not been used in this context before. Eq.~\eqref{eq:hat_sigma_0_moments} together with Eq.~\eqref{eq:eps_moments} immediately imply that: (i) equilibrium densities, (ii) flat-interface profile and (iii) surface tension all depend at the same on the value of $e_4$, hence, it is simply not possible to tune these quantities independently. This point was not completely understood in the seminal works~\cite{Sbragaglia_2007,Falcucci_2007,Chibbaro_2008} and the same conclusion, although using a different perspective, is at the foundations of the results of~\cite{Li_2013} and later~\cite{Khajepor_2015,khajepor2016multipseudopotential}. However, it is important to highlight that in practice, for the settings analyzed in~\cite{Sbragaglia_2007,Falcucci_2007,Chibbaro_2008}, the decoupling was effectively realized withing a reasonable precision.

Such considerations on $\varepsilon$ and $e_4$ are valid when considering Guo's forcing implementation~\cite{Guo_2002} for which all derivative terms can be traced back to the lattice pressure tensor. However, there is another element that has not always fully brought to the attention, namely that the seminal works~\cite{Sbragaglia_2007,Falcucci_2007,Chibbaro_2008} and the later works~\cite{Li_2013,Khajepor_2015,khajepor2016multipseudopotential} have been carried forward using forcing implementations that would, implicitly through the use of the Shan-Chen forcing implementation~\cite{Shan_1993,Shan_1994} for the former and explicitly for the latter, bring into the pressure tensor terms of the form $F^\alpha F^\beta$ which from now on we will refer to as $F^2$ terms. Indeed, it is not possible to obtain the same results as in~~\cite{Sbragaglia_2007,Falcucci_2007,Chibbaro_2008} (in particular the negative surface tension results in~\cite{Chibbaro_2008}) when using Guo's forcing implementation as the $F^2$ terms offer a sort of stabilization mechanism for the interface against the action of the spurious currents. Furthermore, in~\cite{Li_2013,Khajepor_2015,khajepor2016multipseudopotential} it was understood that the $F^2$ terms would also allow for a tuning of $\varepsilon$ that does not depend on the weights of the forcing stencil itself. From the present perspective this can be understood as follows: an $F^2$ term in the pressure tensor would contribute with terms proportional to the square of the pseudopotential derivative $(\mbox{d}\psi / \mbox{d}x)^2$ (for the symmetry reasons discussed at Eq.~\eqref{eq:FSCTaylor} there would be no second derivatives), thus (additively) changing the coefficient $\bar{\alpha}$ in Eq.~\eqref{eq:SCPNExpansion} which in turn would shift the value of $\varepsilon=-2\bar{\alpha}/\bar{\beta}$. Hence, from the perspective of the present work, one could split the value of $\varepsilon$ into a geometric term $\varepsilon_\text{G}$, due to the interaction stencil, and a forcing term $\varepsilon_\text{F}$ due to the specific forcing implementation
\begin{equation}
\varepsilon=\varepsilon_{\text{G}}+\varepsilon_{\text{F}}=\frac{6e_{4}-2e_{2}}{6e_{4}+e_{2}}+\varepsilon_{\text{F}},
\end{equation}
where $\varepsilon_\text{F}=0$ for the present work, i.e., for Guo's forcing implementation~\cite{Guo_2002}. Both lines of research presented in~\cite{Li_2013} and~\cite{khajepor2016multipseudopotential} eventually resorted to an extended Guo's scheme, in~\cite{Li_2013} under the guise of a Multiple-Relaxation-Time (MRT) scheme and in~\cite{khajepor2016multipseudopotential} with the introduction of the $F^2$ terms in the rank-two tensor used in~\eqref{eq:GuoForcing}, namely
\begin{equation}
T_{\alpha\beta} = 2u_\alpha F_\beta - \frac{\mathcal{A}}{\tau} F_\alpha F_\beta,
\end{equation}
with some constant $\mathcal{A}$~\cite{khajepor2016multipseudopotential}, thus yielding a non-vanishing contribution $\varepsilon_{\text{F}}\neq 0$. Indeed, both in~\cite{Li_2013} and~\cite{khajepor2016multipseudopotential} the interaction stencil has been chosen to be the simplest fourth-order isotropic one described above yielding $\varepsilon_\text{G}=0$ so that the prefactor $\mathcal{A}$ could be entirely responsible for tuning the interface thickness through $\varepsilon_{\text{F}}$. On the other hand the prefactor in the expression of the surface tension $c_{02}-c_{11}$ was still found to be proportional to $e_4$ thus effectively decoupling surface tension and density profile.

In the present work, we wish to outline a different strategy, that only relies on the geometrical properties of the Shan-Chen forcing stencil rather than on the modification of the forcing scheme itself. Naturally, this approach should be understood as a complement to the preexisting literature rather than an alternative, as many fundamental points, especially with regards to the implementation of non-trivial equations of state and to the stability of the simulations, have already been drawn in~\cite{Li_2013,Khajepor_2015,khajepor2016multipseudopotential}.

\renewcommand{\arraystretch}{2.6}
\begin{table}[t!]
  \begin{ruledtabular}
  \centering
  \begin{tabular}{lc}
  \multicolumn{2}{c}{One-dimensional Lattice Pressure Tensor coefficients} \\
  \hline
  $\displaystyle a_{\left[-4,0,4\right]}^{\left(\text{N}\right)}$
  & $\displaystyle 2W\left(16\right)$ \\

  $\displaystyle a_{\left[-3,0,3\right]}^{\left(\text{N}\right)}$ 
  & $\displaystyle \frac{3}{2}W\left(9\right)+3W\left(18\right)$ \\

  $\displaystyle a_{\left[-2,0,2\right]}^{\left(\text{N}\right)}$ 
  & $\displaystyle W\left(4\right)+2W\left(5\right)+2W\left(8\right)$ \\

  $\displaystyle a_{\left[-1,0,1\right]}^{\left(\text{N}\right)}$ 
  & $\displaystyle \frac{1}{2}W\left(1\right)+W\left(2\right)+W\left(5\right)$ \\

  $\displaystyle a_{\left[0\right]}^{\left(\text{N}\right)}$ 
  & $\displaystyle 0$ \\

  $\displaystyle b_{\left[-2,2\right]}^{\left(\text{N}\right)}$ 
  & $\displaystyle 4W\left(16\right)$ \\

  $\displaystyle b_{\left[-1,1\right]}^{\left(\text{N}\right)}$ 
  & $\displaystyle 2W\left(4\right)+4W\left(5\right)+4W\left(8\right)$ \\

  $\displaystyle b_{\left[1,3\right]}^{\left(\text{N}\right)}$ 
  & $\displaystyle 4W\left(16\right)$ \\

  $\displaystyle b_{\left[1,2\right]}^{\left(\text{N}\right)}$ 
  & $\displaystyle 3W\left(9\right)+6W\left(18\right)$ \\
  \hline

  $\displaystyle a_{\left[-4,0,4\right]}^{\left(\text{T}\right)}$
  & $\displaystyle 0$ \\

  $\displaystyle a_{\left[-3,0,3\right]}^{\left(\text{T}\right)}$ 
  & $\displaystyle 3W\left(18\right)$ \\

  $\displaystyle a_{\left[-2,0,2\right]}^{\left(\text{T}\right)}$ 
  & $\displaystyle \frac{1}{2}W\left(5\right)+2W\left(8\right)$ \\

  $\displaystyle a_{\left[-1,0,1\right]}^{\left(\text{T}\right)}$ 
  & $\displaystyle W\left(2\right)+4W\left(5\right)$ \\

  $\displaystyle a_{\left[0\right]}^{\left(\text{T}\right)}$ 
  & $\displaystyle W\left(1\right)+4W\left(4\right)+9W\left(9\right)+16W\left(16\right)$\\

  $\displaystyle b_{\left[-2,2\right]}^{\left(\text{T}\right)}$ 
  & $\displaystyle 0$\\

  $\displaystyle b_{\left[-1,1\right]}^{\left(\text{T}\right)}$ 
  & $\displaystyle W\left(5\right)+4W\left(8\right)$\\

  $\displaystyle b_{\left[1,3\right]}^{\left(\text{T}\right)}$ 
  & $\displaystyle 0$\\

  $\displaystyle b_{\left[1,2\right]}^{\left(\text{T}\right)}$ 
  & $\displaystyle 6W\left(18\right)$\\
  \end{tabular}
  \end{ruledtabular}
  \caption{Definitions in terms of the two-dimensional weights of the coefficients appearing in Eq.~\eqref{eq:1dPNT}.}\label{tab:pnptcoeffs}
\end{table}
\section{Defining the stencil}\label{app:stencil}
In this Section we report the information needed to obtain the different sets of weights reported in the Tables~\ref{tab:e4_tuning_6},~\ref{tab:e4e6_tuning_8},~\ref{tab:e4e6e8_tuning_6_2d} and~\ref{tab:e4e6e8_tuning_6_3d}. The first step is to write the weights as a function of the isotropy $\{e_{2n}\}$ and anisotropy $\{I_{2n,k}\}$ coefficients. Given that these coefficients are linear combinations of the weights one simply needs to solve the associated linear system, as long as the chosen expressions are linearly independent. We report in Table~\ref{tab:defe2nI2n} the definition for the isotropy and anisotropy coefficients for the two-dimensional case as well as the anisotropy coefficients for three dimensions for the largest number of weights considered in this work. The expressions related to any smalle stencil can be obtained by setting to zero the appropriate weights. This represents the minimal set of information as the isotropy coefficients in three dimensions can be computed from the two-dimensional expression by means of the expressions
\begin{equation}
\begin{split}
&W_{2}(1)=W_{3}(1)+2W_{3}(2)+2W_{3}(5),\\&W_{2}(2)=W_{3}(2)+2W_{3}(3)+2W_{3}(6),\\&W_{2}(4)=W_{3}(4)+2W_{3}(5)+2W_{3}(8),\\&W_{2}(5)=W_{3}(5)+2W_{3}(6)+2W_{(2,2,1)}\left(9\right),\\&W_{2}(8)=W(8)+2W_{(2,2,1)}(9),\\&W_{2}(9)=W_{3}(9)+2W_{3}(18),\\&W_{2}(16)=W_{3}(16),\\&W_{2}(18)=W_{3}(18),
\end{split}\label{eq:wproj}
\end{equation}
representing the projection of the three-dimensional weights $W_3$ onto the two-dimensional ones $W_2$. However, this is not the case for the anisotropy coefficients since their number changes according to the dimension for every given isotropy order larger than six. Hence the three-dimensional anisotropy coefficients are reported in Table~\ref{tab:defe2nI2n}. We wish to remark here that the expression ``anisotropy coefficients'' strictly applies to the fourth-order case as the higher-order expressions have been computed with the method described in~\cite{Sbragaglia_2007}, i.e., by requiring some ratios of the stencil moments to be isotropic, that we can refer to as \emph{isotropy conditions}. However, when all the isotropy conditions of a given order are set to zero these correspond to a linear combination of the anisotropy coefficients as proved in~\cite{Lulli_2021}. Given the absence of a general parametrization of the anisotropic terms in three dimensions it seems a good choice to use the isotropy conditions~\cite{Sbragaglia_2007} which can be defined independently of the dimension. The formulation of a general parametrization of the anisotropy coefficients will be the subject of a future work. 

\renewcommand{\arraystretch}{2}
\begin{table}[t!]
  \begin{ruledtabular}
  \centering
  \begin{tabular}{p{7mm}p{90mm}}
  \multicolumn{2}{c}{Taylor expansion coefficients for $P_\text{N}(x)-P_\text{T}(x)$} \\
  \hline
  $\displaystyle c_{02}$ & $\displaystyle \frac{1}{2}W(1)+6W(4)+6W(5)+\frac{57}{2}W(9)+88W(16)$ \\
  $\displaystyle c_{11}$ & $\displaystyle -2W(4)-3W(5)-12W(9)-40W(16)$ \\
  $\displaystyle c_{04}$ & $\displaystyle \frac{1}{24}W(1)+\frac{3}{2}W(4)+2W(5)+\frac{115}{8}W(9)+\frac{226}{3}W(16)$ \\
  $\displaystyle c_{13}$ & $\displaystyle -\frac{2}{3}W(4)-W(5)-10W(9)-\frac{184}{3}W(16)$ \\
  $\displaystyle c_{22}$ & $\displaystyle \frac{1}{2}W(4)+\frac{3}{4}W(5)+6W(9)+34W(16)$ \\
  \end{tabular}
  \end{ruledtabular}
  \caption{Definitions in terms of the two-dimensional weights of the coefficients of the Taylor expansion $P_\text{N}(x)-P_\text{T}(x)$ used to define the constants $\hat{\sigma}_0$, $\hat{\sigma}_1$ and $\hat{\delta}_0$.}\label{tab:pn_minus_ptcoeffs}
\end{table}

The next step is to define, in terms of the weights, the coefficients of the Taylor expansion of $P_\text{N}(x)-P_\text{T}(x)$, i.e., the difference between the one-dimensional projections of the normal and tangential components of the lattice pressure tensor. In order to do this one needd to to project to one dimension the expressions for the lattice pressure tensor reported in Appendix~\ref{app:LPT2D} and~\ref{app:LPT3D}, i.e., select the $xx$ and $yy$ components for the normal and tangential directions, respectively, and assume that $\psi$ only depends on the $x$ coordinate. Each component of the pressure tensor can be written in the general form
\begin{equation}
\begin{split}
&P_{\bullet}\left(x\right)=Gc_{s}^{2}a_{\left[-4,0,4\right]}^{\left(\bullet\right)}\psi\left(x\right)\left[\psi\left(x+4\right)+\psi\left(x-4\right)\right]\\&+Gc_{s}^{2}a_{\left[-3,0,3\right]}^{\left(\bullet\right)}\psi\left(x\right)\left[\psi\left(x+3\right)+\psi\left(x-3\right)\right]\\&+Gc_{s}^{2}a_{\left[-2,0,2\right]}^{\left(\bullet\right)}\psi\left(x\right)\left[\psi\left(x+2\right)+\psi\left(x-2\right)\right]\\&+Gc_{s}^{2}a_{\left[-1,0,1\right]}^{\left(\bullet\right)}\psi\left(x\right)\left[\psi\left(x+1\right)+\psi\left(x-1\right)\right]\\&+Gc_{s}^{2}a_{\left[0\right]}^{\left(\bullet\right)}\psi^{2}\left(x\right)\\&+Gc_{s}^{2}b_{\left[-2,2\right]}^{\left(\bullet\right)}\psi\left(x+2\right)\psi\left(x-2\right)\\&+Gc_{s}^{2}b_{\left[-1,1\right]}^{\left(\bullet\right)}\psi\left(x+1\right)\psi\left(x-1\right)\\&+Gc_{s}^{2}b_{\left[1,3\right]}^{\left(\bullet\right)}\left[\psi\left(x+3\right)\psi\left(x-1\right)+\psi\left(x+1\right)\psi\left(x-3\right)\right]\\&+Gc_{s}^{2}b_{\left[1,2\right]}^{\left(\bullet\right)}\left[\psi\left(x+2\right)\psi\left(x-1\right)+\psi\left(x+1\right)\psi\left(x-2\right)\right],
\end{split}\label{eq:1dPNT}
\end{equation}
where the symbol $\bullet$ should be replaced with $\text{N}$ or $\text{T}$ for the normal or tangential component, respectively. The related coefficients are reported in Table~\ref{tab:pnptcoeffs}. The following step is to take the Taylor expansion of $P_\text{N}(x)-P_\text{T}(x)$ where the coefficients are associated with a given combination of derivatives according to the correspondence
\begin{equation}
c_{mn} \leftrightarrow \left(\frac{\mbox{d}^m\psi}{\mbox{d}x^m}\right)\left(\frac{\mbox{d}^n\psi}{\mbox{d}x^n}\right).
\end{equation}
These calculations are performed by the custom-written python modules that can be found in the ``idea.deploy'' project and the GitHub repository associated with this work~\href{https://github.com/lullimat/idea.deploy}{https://github.com/lullimat/idea.deploy}. The expressions of the coefficients $c_{mn}$ in terms of the two-dimensional weights are reported in Table~\ref{tab:pn_minus_ptcoeffs}. Given that all the calculations are performed for the one-dimensional projection, the three-dimensional definition can be obtained by means of the transformation in Eq.~\eqref{eq:wproj}. These results allow to write the weights reported in Tables~\ref{tab:e4_tuning_6},~\ref{tab:e4e6_tuning_8},~\ref{tab:e4e6e8_tuning_6_2d} and~\ref{tab:e4e6e8_tuning_6_3d} after leveraging the definitions of $\hat{\sigma}_0$, $\hat{\sigma}_1$ and $\hat{\delta}_0$ in terms of the coefficients $c_{mn}$ reported in Eq.~\eqref{eq:hatsigma01eI} and~\eqref{eq:hatdelta0eI}.

\section{Lattice Pressure Tensor - $d = 2$}\label{app:LPT2D}
In order to extend the pressure tensor to the case of heterogeneous density distributions with interfaces, one would be tempted to consider the next terms in the Taylor expansion of the SC forcing in Eq.~\eqref{eq:FSCTaylor}, which to second order, could still be regrouped as a divergence of a rank-two tensor. However, it was shown in~\cite{Shan_2008} that such expansion would yield equilibrium estimates for the flat interface that are inconsistent with the simulation results, and that one can define a pressure tensor directly defined \emph{on the lattice} rather than resorting to the Taylor expansion of the forcing. Hence the SC model is equipped also with a~\emph{lattice pressure tensor} which, in the case of Guo's forcing implementation, enjoys the remarkable property of implementing the mechanic equilibrium condition for a flat interface profile at the discrete lattice level, i.e., the constancy of the normal component to the interface $P_{\text{N}}(\mathbf{x})=p_0$. In other words, when crossing a flat interface $P_{\text{N}}$ takes on a constant value on each lattice site within machine precision~\cite{Belardinelli_2015,Lulli_2021,lulli2024metastable}. Unfortunately, it is not possible to provide a concise expression for the SC lattice pressure tensor in the general case, however, it is possible to follow a geometric construction with defined rules for any forcing stencil in two and three dimensions~\cite{Shan_2008,From_2019,Lulli_2021}. However, in some specific cases one can use a concise expression. In two dimensions, this is the case for the groups of squared length $\ell^2=1,2,4,8,9,18$ depicted in Fig.~\ref{fig:stencil}
\begin{equation}
\begin{split} & P_{\left\{ 1,2\right\} }^{\mu\nu}\left(\mathbf{x}\right)=\frac{Gc_{s}^{2}}{2}\psi\left(\mathbf{x}\right)\langle\psi\left(\mathbf{x}+\mathbf{e}\right)e^{\mu}e^{\nu}\rangle_{1,2},\\
 & P_{\left\{ 4,8\right\} }^{\mu\nu}\left(\mathbf{x}\right)=\frac{Gc_{s}^{2}}{4}\psi\left(\mathbf{x}\right)\langle\psi\left(\mathbf{x}+\mathbf{e}\right)e^{\mu}e^{\nu}\rangle_{4,8}\\
 & \qquad+\frac{Gc_{s}^{2}}{4}\langle\psi\left(\mathbf{x}+\frac{\mathbf{e}}{2}\right)\psi\left(\mathbf{x}-\frac{\mathbf{e}}{2}\right)e^{\mu}e^{\nu}\rangle_{4,8},\\
\end{split}
\end{equation}
\begin{equation}
\begin{split}
 & P_{\left\{ 9,18\right\} }^{\mu\nu}\left(\mathbf{x}\right)=\frac{Gc_{s}^{2}}{6}\psi\left(\mathbf{x}\right)\langle\psi\left(\mathbf{x}+\mathbf{e}\right)e^{\mu}e^{\nu}\rangle_{9,18}\\
 & \qquad+\frac{Gc_{s}^{2}}{3}\langle\psi\left(\mathbf{x}+\frac{\mathbf{e}}{3}\right)\psi\left(\mathbf{x}-\frac{2\mathbf{e}}{3}\right)e^{\mu}e^{\nu}\rangle_{9,18},
\end{split}
\end{equation}
where the indices near the angled brackets indicate which subset of the weights is used to compute the expected value. When considering the group of squared length $\ell^2=5$ one needs to distinguish two kinds of contributions, namely a regular one
\begin{equation}
P_{\left\{ 5\right\} a}^{\mu\nu}\left(\mathbf{x}\right)=\frac{Gc_{s}^{2}}{4}\psi\left(\mathbf{x}\right)\langle\psi\left(\mathbf{x}+\mathbf{e}\right)e^{\mu}e^{\nu}\rangle_{5},
\end{equation}
and an irregular one
\begin{equation}
\begin{split}
&P_{\left\{ 5\right\} b}^{\mu\nu}\left(\mathbf{x}\right)=\frac{Gc_{s}^{2}}{4}W\left(5\right)\left[\psi_{-1,0}\psi_{1,1}+\psi_{-1,-1}\psi_{1,0}\right]e_{13}^{\mu}e_{13}^{\nu}\\&+\frac{Gc_{s}^{2}}{4}W\left(5\right)\left[\psi_{-1,-1}\psi_{0,1}+\psi_{0,-1}\psi_{1,1}\right]e_{14}^{\mu}e_{14}^{\nu}\\&+\frac{Gc_{s}^{2}}{4}W\left(5\right)\left[\psi_{0,-1}\psi_{-1,1}+\psi_{1,-1}\psi_{0,1}\right]e_{15}^{\mu}e_{15}^{\nu}\\&+\frac{Gc_{s}^{2}}{4}W\left(5\right)\left[\psi_{1,-1}\psi_{-1,0}+\psi_{1,0}\psi_{-1,1}\right]e_{16}^{\mu}e_{16}^{\nu},
\end{split}
\end{equation}
where the shorthand notation $\psi_{a,b} = \psi(\mathbf{x} + a\mathbf{e}_1 + b\mathbf{e}_2)$ is used, with $\mathbf{e}_1$ and $\mathbf{e}_2$ being the basis vector of the $x$ and $y$ directions respectively.
One can check that, when considering all the contributions from the different groups, at leading order one recovers the expectation value for the covariance over the entire stencil, i.e., the usual bulk contribution
\begin{equation}
P^{\mu\nu}\left(\mathbf{x}\right)=\frac{Gc_{s}^{2}}{2}\psi^{2}\left(\mathbf{x}\right)\langle e^{\mu}e^{\nu}\rangle_{1,2,4,5,8,9,18}=\frac{Gc_{s}^{2}e_{2}}{2}\psi^{2}\left(\mathbf{x}\right)\delta^{\mu\nu}.
\end{equation}
From now on we will indicate with $P^{\mu\nu}\left(\mathbf{x}\right)$ the sum of all the contributions from the different symmetry groups to the lattice pressure tensor.
\vspace{2cm}

\begin{widetext}
\section{Lattice Pressure Tensor - $d = 3$}\label{app:LPT3D}
First, we report the fully symmetric groups
\begin{equation}
\begin{split} & P_{\left\{ 1,2,3\right\} }^{\mu\nu}\left(\mathbf{x}\right)=\frac{Gc_{s}^{2}}{2}\psi\left(\mathbf{x}\right)\langle\psi\left(\mathbf{x}+\mathbf{e}\right)e^{\mu}e^{\nu}\rangle_{1,2,3}\\
 & P_{\left\{ 4,8\right\} }^{\mu\nu}\left(\mathbf{x}\right)=\frac{Gc_{s}^{2}}{4}\psi\left(\mathbf{x}\right)\langle\psi\left(\mathbf{x}+\mathbf{e}\right)e^{\mu}e^{\nu}\rangle_{4,8}+\frac{Gc_{s}^{2}}{4}\langle\psi\left(\mathbf{x}+\frac{\mathbf{e}}{2}\right)\psi\left(\mathbf{x}-\frac{\mathbf{e}}{2}\right)e^{\mu}e^{\nu}\rangle_{4,8}\\
 & P_{\left\{ 9,18\right\} }^{\mu\nu}\left(\mathbf{x}\right)=\frac{Gc_{s}^{2}}{6}\psi\left(\mathbf{x}\right)\langle\psi\left(\mathbf{x}+\mathbf{e}\right)e^{\mu}e^{\nu}\rangle_{9,18}+\frac{Gc_{s}^{2}}{3}\langle\psi\left(\mathbf{x}+\frac{\mathbf{e}}{3}\right)\psi\left(\mathbf{x}-\frac{2\mathbf{e}}{3}\right)e^{\mu}e^{\nu}\rangle_{9,18}\\
 & P_{\left\{ 16\right\} }^{\mu\nu}\left(\mathbf{x}\right)=\frac{Gc_{s}^{2}}{8}\psi\left(\mathbf{x}\right)\langle\psi\left(\mathbf{x}+\mathbf{e}_{a}\right)e^{\mu}e^{\nu}\rangle_{16}+\frac{Gc_{s}^{2}}{4}\langle\psi\left(\mathbf{x}+\frac{\mathbf{e}}{4}\right)\psi\left(\mathbf{x}-\frac{3\mathbf{e}}{4}\right)e^{\mu}e^{\nu}\rangle_{16}+\frac{Gc_{s}^{2}}{8}\langle\psi\left(\mathbf{x}+\frac{\mathbf{e}}{2}\right)\psi\left(\mathbf{x}-\frac{\mathbf{e}}{2}\right)e^{\mu}e^{\nu}\rangle_{16}.
\end{split}
\end{equation}
Then, the regular parts of the irregular groups
\begin{equation}
\begin{split}P_{(2,1,0),(2,1,1),(2,2,1)a}^{\mu\nu}\left(\mathbf{x}\right)= & \frac{Gc_{s}^{2}}{4}\psi\left(\mathbf{x}\right)\langle\psi\left(\mathbf{x}+\mathbf{e}\right)e^{\mu}e^{\nu}\rangle_{(2,1,0),(2,1,1),(2,2,1)},\end{split}
\end{equation}
and finally the irregular contributions
\begin{equation}
\begin{split} & P_{\left[2,1,0\right]b}^{\mu\nu}\left(\mathbf{x}\right)=\frac{Gc_{s}^{2}}{4}W\left(5\right)\left\{ \left[\psi_{-1,0,0}\psi_{1,1,0}+\psi_{-1,-1,0}\psi_{1,0,0}\right]e_{\left(2,1,0\right)}^{\mu}e_{\left(2,1,0\right)}^{\nu}+\left[\psi_{-1,-1,0}\psi_{0,1,0}+\psi_{0,-1,0}\psi_{1,1,0}\right]e_{\left(1,2,0\right)}^{\mu}e_{\left(1,2,0\right)}^{\nu}\right\} \\
 & +\frac{Gc_{s}^{2}}{4}W\left(5\right)\left\{ \left[\psi_{0,-1,0}\psi_{-1,1,0}+\psi_{1,-1,0}\psi_{0,1,0}\right]e_{\left(-1,2,0\right)}^{\mu}e_{\left(-1,2,0\right)}^{\nu}+\left[\psi_{1,-1,0}\psi_{-1,0,0}+\psi_{1,0,0}\psi_{-1,1,0}\right]e_{\left(-2,1,0\right)}^{\mu}e_{\left(-2,1,0\right)}^{\nu}\right\} \\
\\ & +\frac{Gc_{s}^{2}}{4}W\left(5\right)\left\{ \left[\psi_{0,-1,0}\psi_{0,1,1}+\psi_{0,-1,-1}\psi_{0,1,0}\right]e_{\left(0,2,1\right)}^{\mu}e_{\left(0,2,1\right)}^{\nu}+\left[\psi_{0,-1,-1}\psi_{0,0,1}+\psi_{0,0,-1}\psi_{0,1,1}\right]e_{\left(0,1,2\right)}^{\mu}e_{\left(0,1,2\right)}^{\nu}\right\} \\
 & +\frac{Gc_{s}^{2}}{4}W\left(5\right)\left\{ \left[\psi_{0,0,-1}\psi_{0,-1,1}+\psi_{0,1,-1}\psi_{0,0,1}\right]e_{\left(0,-1,2\right)}^{\mu}e_{\left(0,-1,2\right)}^{\nu}+\left[\psi_{0,1,-1}\psi_{0,-1,0}+\psi_{0,1,0}\psi_{0,-1,1}\right]e_{\left(0,-2,1\right)}^{\mu}e_{\left(0,-2,1\right)}^{\nu}\right\} \\
\\ & +\frac{Gc_{s}^{2}}{4}W\left(5\right)\left\{ \left[\psi_{0,0,-1}\psi_{1,0,1}+\psi_{-1,0,-1}\psi_{0,0,1}\right]e_{\left(1,0,2\right)}^{\mu}e_{\left(1,0,2\right)}^{\nu}+\left[\psi_{-1,0,-1}\psi_{1,0,0}+\psi_{-1,0,0}\psi_{1,0,1}\right]e_{\left(2,0,1\right)}^{\mu}e_{\left(2,0,1\right)}^{\nu}\right\} \\
 & +\frac{Gc_{s}^{2}}{4}W\left(5\right)\left\{ \left[\psi_{-1,0,0}\psi_{1,0,-1}+\psi_{-1,0,1}\psi_{1,0,0}\right]e_{\left(2,0,-1\right)}^{\mu}e_{\left(2,0,-1\right)}^{\nu}+\left[\psi_{-1,0,1}\psi_{0,0,-1}+\psi_{0,0,1}\psi_{1,0,-1}\right]e_{\left(1,0,-2\right)}^{\mu}e_{\left(1,0,-2\right)}^{\nu}\right\} 
\end{split}
\end{equation}
\begin{equation}
\begin{split} & P_{\left[2,1,1\right]b}^{\mu\nu}\left(\mathbf{x}\right)=\frac{Gc_{s}^{2}}{8}W\left(6\right)\left[\psi_{-1,0,0}\psi_{1,1,1}+\psi_{-1,0,-1}\psi_{1,1,0}+\psi_{-1,-1,0}\psi_{1,0,1}+\psi_{-1,-1,-1}\psi_{1,0,0}\right]e_{\left(2,1,1\right)}^{\mu}e_{\left(2,1,1\right)}^{\nu}\\
 & +\frac{Gc_{s}^{2}}{8}W\left(6\right)\left[\psi_{-1,-1,0}\psi_{0,1,1}+\psi_{-1,-1,-1}\psi_{0,1,0}+\psi_{0,-1,0}\psi_{1,1,1}+\psi_{0,-1,-1}\psi_{1,1,0}\right]e_{\left(1,2,1\right)}^{\mu}e_{\left(1,2,1\right)}^{\nu}\\
 & +\frac{Gc_{s}^{2}}{8}W\left(6\right)\left[\psi_{0,-1,0}\psi_{-1,1,1}+\psi_{0,-1,-1}\psi_{-1,1,0}+\psi_{1,-1,0}\psi_{0,1,1}+\psi_{1,-1,-1}\psi_{0,1,0}\right]e_{\left(-1,2,1\right)}^{\mu}e_{\left(-1,2,1\right)}^{\nu}\\
 & +\frac{Gc_{s}^{2}}{8}W\left(6\right)\left[\psi_{1,-1,0}\psi_{-1,0,1}+\psi_{1,-1,-1}\psi_{-1,0,0}+\psi_{1,0,0}\psi_{-1,1,1}+\psi_{1,0,-1}\psi_{-1,1,0}\right]e_{\left(-2,1,1\right)}^{\mu}e_{\left(-2,1,1\right)}^{\nu}\\
\\ 
& +\frac{Gc_{s}^{2}}{8}W\left(6\right)\left[\psi_{-1,-1,1}\psi_{1,0,0}+\psi_{-1,-1,0}\psi_{1,0,-1}+\psi_{-1,0,1}\psi_{1,1,0}+\psi_{-1,0,0}\psi_{1,1,-1}\right]e_{\left(2,1,-1\right)}^{\mu}e_{\left(2,1,-1\right)}^{\nu}\\
 & +\frac{Gc_{s}^{2}}{8}W\left(6\right)\left[\psi_{-1,-1,1}\psi_{0,1,0}+\psi_{-1,-1,0}\psi_{0,1,-1}+\psi_{0,-1,1}\psi_{1,1,0}+\psi_{0,-1,0}\psi_{1,1,-1}\right]e_{\left(1,2,-1\right)}^{\mu}e_{\left(1,2,-1\right)}^{\nu}\\
 & +\frac{Gc_{s}^{2}}{8}W\left(6\right)\left[\psi_{0,-1,1}\psi_{-1,1,0}+\psi_{0,-1,0}\psi_{-1,1,-1}+\psi_{1,-1,1}\psi_{0,1,0}+\psi_{1,-1,0}\psi_{0,1,-1}\right]e_{\left(-1,2,-1\right)}^{\mu}e_{\left(-1,2,-1\right)}^{\nu}\\
 & +\frac{Gc_{s}^{2}}{8}W\left(6\right)\left[\psi_{1,-1,1}\psi_{-1,0,0}+\psi_{1,-1,0}\psi_{-1,0,-1}+\psi_{1,0,1}\psi_{-1,1,0}+\psi_{1,0,0}\psi_{-1,1,-1}\right]e_{\left(-2,1,-1\right)}^{\mu}e_{\left(-2,1,-1\right)}^{\nu}\\
\\ & +\frac{Gc_{s}^{2}}{8}W\left(6\right)\left[\psi_{-1,-1,-1}\psi_{0,0,1}+\psi_{0,-1,-1}\psi_{1,0,1}+\psi_{-1,0,-1}\psi_{0,1,1}+\psi_{0,0,-1}\psi_{1,1,1}\right]e_{\left(1,1,2\right)}^{\mu}e_{\left(1,1,2\right)}^{\nu}\\
 & +\frac{Gc_{s}^{2}}{8}W\left(6\right)\left[\psi_{0,-1,-1}\psi_{-1,0,1}+\psi_{1,-1,-1}\psi_{0,0,1}+\psi_{0,0,-1}\psi_{-1,1,1}+\psi_{1,0,-1}\psi_{0,1,1}\right]e_{\left(-1,1,2\right)}^{\mu}e_{\left(-1,1,2\right)}^{\nu}\\
 & +\frac{Gc_{s}^{2}}{8}W\left(6\right)\left[\psi_{-1,-1,1}\psi_{0,0,-1}+\psi_{0,-1,1}\psi_{1,0,-1}+\psi_{0,0,1}\psi_{1,1,-1}+\psi_{-1,0,1}\psi_{0,1,-1}\right]e_{\left(1,1,-2\right)}^{\mu}e_{\left(1,1,-2\right)}^{\nu}\\
 & +\frac{Gc_{s}^{2}}{8}W\left(6\right)\left[\psi_{0,-1,1}\psi_{-1,0,-1}+\psi_{1,-1,1}\psi_{0,0,-1}+\psi_{1,0,1}\psi_{0,1,-1}+\psi_{0,0,1}\psi_{-1,1,-1}\right]e_{\left(-1,1,-2\right)}^{\mu}e_{\left(-1,1,-2\right)}^{\nu}
\end{split}
\end{equation}

\begin{equation}
\begin{split}P_{\left[2,2,1\right]b}^{\mu\nu}\left(\mathbf{x}\right)= & \frac{Gc_{s}^{2}}{4}W_{\left[2,2,1\right]}\left[\psi_{-1,-1,-1}\psi_{1,1,0}+\psi_{-1,-1,0}\psi_{1,1,1}\right]e_{\left(2,2,1\right)}^{\mu}e_{\left(2,2,1\right)}^{\nu}\\
+ & \frac{Gc_{s}^{2}}{4}W_{\left[2,2,1\right]}\left[\psi_{1,-1,-1}\psi_{-1,1,0}+\psi_{1,-1,0}\psi_{-1,1,1}\right]e_{\left(-2,2,1\right)}^{\mu}e_{\left(-2,2,1\right)}^{\nu}\\
+ & \frac{Gc_{s}^{2}}{4}W_{\left[2,2,1\right]}\left[\psi_{-1,-1,0}\psi_{1,1,-1}+\psi_{-1,-1,1}\psi_{-1,1,0}\right]e_{\left(2,2,-1\right)}^{\mu}e_{\left(2,2,-1\right)}^{\nu}\\
+ & \frac{Gc_{s}^{2}}{4}W_{\left[2,2,1\right]}\left[\psi_{1,-1,0}\psi_{-1,1,-1}+\psi_{1,-1,1}\psi_{-1,1,0}\right]e_{\left(-2,2,-1\right)}^{\mu}e_{\left(-2,2,-1\right)}^{\nu}\\
\\+ & \frac{Gc_{s}^{2}}{4}W_{\left[2,2,1\right]}\left[\psi_{-1,-1,-1}\psi_{1,0,1}+\psi_{-1,0,-1}\psi_{1,1,1}\right]e_{\left(2,1,2\right)}^{\mu}e_{\left(2,1,2\right)}^{\nu}\\
+ & \frac{Gc_{s}^{2}}{4}W_{\left[2,2,1\right]}\left[\psi_{-1,-1,-1}\psi_{0,1,1}+\psi_{0,-1,-1}\psi_{1,1,1}\right]e_{\left(1,2,2\right)}^{\mu}e_{\left(1,2,2\right)}^{\nu}\\
+ & \frac{Gc_{s}^{2}}{4}W_{\left[2,2,1\right]}\left[\psi_{0,-1,-1}\psi_{-1,1,1}+\psi_{1,-1,-1}\psi_{0,1,1}\right]e_{\left(-1,2,2\right)}^{\mu}e_{\left(-1,2,2\right)}^{\nu}\\
+ & \frac{Gc_{s}^{2}}{4}W_{\left[2,2,1\right]}\left[\psi_{1,-1,-1}\psi_{-1,0,1}+\psi_{1,0,-1}\psi_{-1,1,1}\right]e_{\left(-2,1,2\right)}^{\mu}e_{\left(-2,1,2\right)}^{\nu}\\
\\+ & \frac{Gc_{s}^{2}}{4}W_{\left[2,2,1\right]}\left[\psi_{-1,-1,1}\psi_{1,0,-1}+\psi_{-1,0,1}\psi_{1,1,-1}\right]e_{\left(2,1,-2\right)}^{\mu}e_{\left(2,1,-2\right)}^{\nu}\\
+ & \frac{Gc_{s}^{2}}{4}W_{\left[2,2,1\right]}\left[\psi_{-1,-1,1}\psi_{0,1,-1}+\psi_{0,-1,1}\psi_{1,1,-1}\right]e_{\left(1,2,-2\right)}^{\mu}e_{\left(1,2,-2\right)}^{\nu}\\
+ & \frac{Gc_{s}^{2}}{4}W_{\left[2,2,1\right]}\left[\psi_{0,-1,1}\psi_{-1,1,-1}+\psi_{1,-1,1}\psi_{0,1,-1}\right]e_{\left(-1,2,-2\right)}^{\mu}e_{\left(-1,2,-2\right)}^{\nu}\\
+ & \frac{Gc_{s}^{2}}{4}W_{\left[2,2,1\right]}\left[\psi_{1,-1,1}\psi_{-1,0,-1}+\psi_{1,0,1}\psi_{-1,1,-1}\right]e_{\left(-2,1,-2\right)}^{\mu}e_{\left(-2,1,-2\right)}^{\nu}
\end{split}
\end{equation}
\end{widetext}

\bibliographystyle{apsrev4-2}
\bibliography{common-biblio}

\label{lastpage}
\end{document}